\documentclass[11pt]{article} 
\usepackage{amsmath,amssymb} 
\usepackage[dvips]{graphicx,color}

\numberwithin{equation}{section}

\begin{document}

\title{Schr\"{o}dinger-Cat-Likeness 
in Adiabatic Approximation for Generalized Quantum Rabi Model 
without and with $A^{2}$-Term}

\author{Masao Hirokawa \\ 
{\footnotesize Institute of Engineering, Hiroshima University, Higashi-Hiroshima, 739-8527, Japan} 
}

\date{}
\maketitle

\begin{abstract}
We give a mathematical procedure to obtain 
the adiabatic approximation for the generalized 
quantum Rabi Hamiltonian both without and with 
a quadratic interaction. 
We consider the Hamiltonian as the energy of a model 
describing the interaction system of a two-level artificial atom 
and a one-mode microwave photon in circuit QED. 
In the case without the quadratic interaction, 
we show in the adiabatic approximation 
that whether each bare eigenstate forms a Schr\"{o}dinger-cat-like 
entangled state or not depends on whether the energy bias 
of the atom is zero or non-zero, 
and then, the effect of the tunnel splitting 
of the atom is ignored. 
On the other hand, in the case with the quadratic interaction, 
we show in the adiabatic approximation 
that all the physical eigenstates obtained by 
the (meson) pair theory 
form individual Schr\"{o}dinger-cat-like entangled states 
for every energy bias. 
We conclude that this fact comes from the effect of the tunnel splitting. 
\end{abstract}

\maketitle


\section{Introduction}
\label{sec:intro}

Quantum electrodynamics (QED) describes the interaction 
between light and matter. 
QED is a great success as a quantum field theory (QFT) 
for electrodynamics, 
which certifies that QFT is useful and excellent 
to explain the electromagnetic force 
caused by the exchanging of photons. 
The exchanged photon is called a virtual photon 
\cite{PRSZ}. 
QED enables us predict some quantities 
such as the Lamb shift, the difference in energy between 
the two energy levels of the two orbitals 
$2S_{1/2}$ and $2P_{1/2}$, 
of hydrogen atom with extreme accuracy 
\cite{feynman,BD1,BD2,cohen-tannoudji,kaku,PS}. 
The Lamb shift is caused by the fact that 
the different orbitals interact with 
the vacuum fluctuations of the radiation field. 
The vacuum fluctuation is originated from 
the annihilation and creation of virtual photons; 
therefore, the Lamb shift results from 
the fact that the atom is dressed 
with the cloud of virtual photons. 
Even the ground state is a non-zero photon 
state; however, the photons with which it is dressed 
are virtual and not directly observed  
(cf. Complement B${}_{\mbox{\tiny III}}$.2 of Ref.\cite{cohen-tannoudji}). 
It is well known that the vacuum fluctuation due to 
Heisenberg's uncertainty principle brings 
the generation of virtual particles from the quantum vacuum 
\cite{nori12}. 
The state dressed with the virtual photons 
is called the bare state. 
Some method have been considered to derive physical states, 
which are experimentally observable states, 
from the bare states \cite{cohen-tannoudji,HT}.  
After the success of QED, some physicists 
developed the analogy for QED, 
and applied it to nuclear models. 
They then had to meet and straggle troubles of the strong interaction. 
Following Yukawa's theory \cite{yukawa}, 
nucleons are connected by a strong force, 
called nuclear force, and it is made by the fact 
that nucleons exchange $\pi$-mesons (i.e., pion).    
Namely, nucleon and $\pi$-meson respectively play individual roles 
of electron and photon in QED. 
In the early 1940s, (meson) pair theory were studied by Wentzel 
\cite{wentzel1,wentzel2} 
to consider the nuclear forces under the strong coupling regime 
\cite{BR}.  
On another note, according to quantum cromodynamics (QCD), 
quark and gluon in QCD respectively play roles 
of electron and photon in QED. 
Hadrons are classified into mesons and baryons 
consisting of quarks. 
Thus, the well-known problem that whether 
we can derive Yukawa's theory for the nuclear force 
from QCD arises. 

The recent technology of circuit QED can make a quantum simulation 
of cavity QED. 
Quantum simulation is to simulate a target quantum system 
by a controllable quantum system \cite{BN}. 
In particular, it enables us experimentally 
to demonstrate the amazingly strong interaction between 
a two-level artificial atom and a one-mode light on a superconducting circuit: 
Cavity QED has supplied us with 
stronger interaction than the standard QED does \cite{HR-RBH,rbh01}.  
Experimental physicists demonstrate the interaction 
using a two-level atom coupled with a one-mode light in a mirror cavity. 
The solid-state analogue of the strong interaction 
in a superconducting system was theoretically proposed \cite{MSS,MB}, 
and it has been experimentally demonstrated \cite{Chiorescu, Wallraff08,Wallraff04}. 
The atom, the light, and the mirror resonator in cavity QED 
are respectively replaced by an artificial atom, 
a microwave, and a microwave resonator 
on a superconducting circuit. 
The artificial atom is a superconducting LC circuit 
based on some Josephson junctions. 
This replaced cavity QED is the so-called circuit QED 
\cite{YN,nori17}. 
The circuit QED has been intensifying the coupling strength 
so that its region is beyond the strong coupling regime 
\cite{DGS,ciuti,Mooij,Gross,FD,yoshihara1}.

Yoshihara \textit{et al}. succeeded in demonstrating 
the deep-strong coupling regime, 
and experimentally showed how the theory using 
the quantum Rabi model can well describe 
a physical set-up of circuit QED 
\cite{yoshihara1,yoshihara2,yoshihara3}. 
The set-up consists of a two-level artificial atom interacting to 
a one-mode photon of a microwave cavity. 
The notion of the deep-strong coupling regime is 
proposed in Ref.\cite{CS}, 
and the strength of that regime 
is so large that it exceeds the strength of 
the ultra-strong coupling regime 
for the atom-photon interaction in circuit QED. 
Braak gives an analytics solution of the eigenvalue problem 
for the quantum Rabi model \cite{braak1}. 
In Ref.\cite{AN}, meanwhile, 
Ashhab and Nori give a physical establishment of 
the adiabatic approximation \cite{irish} 
for the bare eigenstates of 
the quantum Rabi model. 
The adiabatically approximated eigenstates make 
the Schr\"{o}dinger-cat-like entangled states. 
The adiabatic approximation is very handy to analyze the quantum Rabi model, 
and thus, 
the Schr\"{o}dinger-cat-likeness is beginning to investigate \cite{fuse} 
using it. 
In Ref.\cite{HMS}, we show a mathematical theory 
so that the adiabatic approximation is actually obtained 
under the strong-coupling limit 
in the norm resolvent sense. 

We are interested in a quantum simulation 
of some phenomena predicted in nuclear physics on superconducting circuit. 
In particular, this paper deals with the (meson) pair theory for 
the generalized quantum Rabi Hamiltonian, 
which is also called asymmetric quantum Rabi Hamiltonian. 
It consists of the two-level atom Hamiltonian, the one-mode photon Hamiltonian, 
and the interaction between the atom and the photon. 
We give our attention to the non-zero energy bias 
in the atom Hamiltonian. 
In the case where the energy bias is equal to zero, 
the generalized quantum Rabi model is the quantum Rabi model. 
The energy-bias parameter is easily tunable in 
experiments of circuit QED with the cutting-edge technology. 
Thus, we treat it as a tunable parameter. 
We consider the bare (physical) eigenstates 
of the generalized quantum Rabi Hamiltonian 
without (with) the quadratic interaction. 
The quadratic interaction is often called 
the $A^{2}$-term. 
We then show how we can mathematically obtain 
the adiabatic approximation for 
the generalized quantum Rabi Hamiltonian 
both without and with the $A^{2}$-term. 
Based on this mathematical theory, 
in the case without the $A^{2}$-term, 
we show that whether the adiabatically approximated bare eigenstates 
are formed as the Schr\"{o}dinger-cat-like entangled states 
or not depends on whether the energy bias 
is zero or non-zero. 
As its result, we point out that the effect of the tunnel splitting 
of the two-level atom disappears. 
On the other hand, in the case with the $A^{2}$-term, 
we renormalize it using (meson) pair theory 
\cite{wentzel1,wentzel2,BR}, 
and show that all the adiabatically approximated 
physical eigenstates are formed as the Schr\"{o}dinger-cat-like 
entangled states for every energy bias. 
We realize that this fact results from the effect of 
the tunnel splitting of the two-level atom.

\section{Mathematical Set-Ups}

In this section, we prepare and recall some mathematical 
notations and notions 
to explain and consider our problem. 
For their details, see Refs.\cite{RS1, weidmann} 
for instance. 

For a separable Hilbert space $\mathfrak{H}$ 
we denote its inner product by $(\,\,\, , \,\,\,)_{\mathfrak{H}}$.  
The norm $\|\,\,\, \|_{\mathfrak{H}}$ is naturally introduced 
by $\|\psi\|_{\mathfrak{H}}
=\sqrt{(\psi , \psi)_{\mathfrak{H}}}$ for every 
vector $\psi$ in the Hilbert space $\mathfrak{H}$. 
An operator $A$ acting in the Hilbert space 
is the linear map from a linear subspace $D(A)\subset\mathfrak{H}$ 
to the Hilbert space $\mathfrak{H}$. 
The subspace $D(A)$ is called the \textit{domain} of the operator $A$. 
In particular, when the operator $A$ satisfies that 
there is a positive constant $M$ so that  
the inequality, $\|A\psi\|_{\mathfrak{H}}
\le M\|\psi\|_{\mathfrak{H}}$, can hold 
for any vector $\psi\in \mathfrak{H}$, 
we say the operator $A$ is \textit{bounded}. 
Then, for every bounded operator $A$, 
the operator norm is given by 
$\| A\|_{\mathrm{op}}:=
\sup_{\psi\in D(A)\,\mathrm{with}\,\psi\ne 0}
\| A\psi\|_{\mathfrak{H}}/\|\psi\|_{\mathfrak{H}}$. 
The inequality, $\| A\psi\|_{\mathfrak{H}}
\le \|A\|_{\mathrm{op}}\|\psi\|_{\mathfrak{H}}$, holds then.  
On the other hand, in the case $\| A\|_{\mathrm{op}}=\infty$, 
we say the operator $A$ is \textit{unbounded}. 
In quantum theory, an observable $A$ corresponds to 
a self-adjoint operator, that is, 
it satisfies the domain identity, $D(A)=D(A^{*})$, 
and the action identity, $A\psi=A^{*}\psi$ 
for every vector $\psi\in D(A)$, 
where $A^{*}$ is the adjoint operator of the operator $A$. 
It is convenient to consider the resolvent $(A-iz)^{-1}$ 
for an unbounded self-adjoint operator $A$ 
for every complex number $z$ with $\Im z\ne 0$. 
Let $A_{n}$ be a sequence of self-adjoint operators. 
When there is a self-adjoint operator $A$ so that 
the limit, $\lim_{n\to\infty}
\|(A_{n}-iz)^{-1}-(A-iz)^{-1}\|_{\mathrm{op}}=0$, 
holds for every complex number $z$ with $\Im z\ne 0$, 
the operators $A_{n}$ are said to converge 
to the operator $A$ in the \textit{norm resolvent sense} 
\cite{RS1}, 
and we often denote the convergence by 
$A_{n} 
\stackrel{\mathrm{n.r.s.}\,\,}{{-\!\!-\!\!\!}\longrightarrow} 
A
\,\,\, \mbox{as $n\to\infty$}
$ 
in this paper.

We sometimes represent by $|E\rangle$ a vector in 
the state space $\mathfrak{H}$. 
We often use Dirac's bra-ket  notation 
$\langle E_{1}|E_{2}\rangle$ 
for the inner product 
$(|E_{2}\rangle , |E_{2}\rangle)_{\mathfrak{H}}$ 
of vectors $|E_{1}\rangle$ and $|E_{2}\rangle$, i.e., 
$\langle E_{1}|E_{2}\rangle:=
(|E_{2}\rangle , |E_{2}\rangle)_{\mathfrak{H}}$. 
So, the notation $\langle E_{1}|A|E_{2}\rangle$ 
stands for the inner product 
$(|E_{2}\rangle , A|E_{2}\rangle)_{\mathfrak{H}}$ 
for vectors $|E_{1}\rangle$, $|E_{2}\rangle$ and an operator $A$, 
i.e., $\langle E_{1}|A|E_{2}\rangle:=
(|E_{2}\rangle , A|E_{2}\rangle)_{\mathfrak{H}}$. 

Let $A_{j}$ be an operator acting in Hilbert spaces $\mathfrak{H}$, 
$j=1, 2$. 
We often omit the tensor sign $\otimes$ from 
the tensor product $A_{1}\otimes A_{2}$, 
and denote the tensor product $A_{1}\otimes A_{2}$ by $A_{1}A_{2}$, 
i.e., $A_{1}A_{2}=A_{1}\otimes A_{2}$.  
We simply write $A_{1}\otimes I_{\mathfrak{H}_{2}}$ as $A_{1}$ 
for the identity operator $I_{\mathfrak{H}_{2}}$, 
and $I_{\mathfrak{H}_{1}}\otimes A_{2}$ as $A_{2}$ 
for the identity matrix $I_{\mathfrak{H}_{1}}$.  
Correspondingly, we also omit the tensor symbol $\otimes$ form 
the tensor product of vectors in $\mathfrak{H}_{1}
\otimes\mathfrak{H}_{2}$.

The state space of the two-level atom system coupled 
with one-mode light is given by $\mathbb{C}^{2}\otimes L^{2}(\mathbb{R})$, 
where $\mathbb{C}^{2}$ is the $2$-dimensional unitary space, 
and $L^{2}(\mathbb{R})$ the Hilbert space consisting of 
the square-integrable functions. 
We sometimes omit the tensor sign $\otimes$ from 
vectors in the state space 
$\mathbb{C}^{2}\otimes L^{2}(\mathbb{R})$. 
We use the notation $|\!\!\uparrow\rangle$ 
for the up-spin state
and the notation $|\!\!\downarrow\rangle$ 
for the down-spin state, 
which are defined by 
$
|\!\!\uparrow\rangle
:=
{\scriptsize 
\left(\hspace*{-1.5mm}
\begin{array}{cc}
1 \\ 
0 
\end{array}
\hspace*{-1.5mm}
\right)
}$
and 
$ 
|\!\!\downarrow\rangle
:=
{\scriptsize 
\left(\hspace*{-1.5mm}
\begin{array}{cc}
0 \\ 
1 
\end{array}
\hspace*{-1.5mm}
\right)
}$ 
in $\mathbb{C}^{2}$. 
We use standard notations for the Pauli matrices, 
$\sigma_{x}:=
{\scriptsize 
\left(\hspace*{-1.5mm}
\begin{array}{cc}
0 & 1 \\ 
1 & 0
\end{array}
\hspace*{-1.5mm}\right)
}$, 
$\sigma_{y}:=
{\scriptsize 
\left(\hspace*{-1.5mm}
\begin{array}{cc}
0 & -i \\ 
i & 0
\end{array}
\hspace*{-1.5mm}\right)
}$, 
and  
$\sigma_{z}:=
{\scriptsize 
\left(\hspace*{-1.5mm}
\begin{array}{cc}
1 & 0 \\ 
0 & -1
\end{array}
\hspace*{-1.5mm}
\right)
}$. 
We denote by $|n\rangle$ the Fock state in $L^{2}(\mathbb{R})$ 
with the photon number 
$n=0, 1, 2, \cdots$. 
That is, $|0\rangle:=(w^{2}/\pi)^{1/4}
\exp\left[-(wx)^{2}/2\right]$ and 
$|n\rangle:=\sqrt{w}\gamma_{n}H_{n}(wx)\exp\left[-(wx)^{2}/2\right] 
\in L^{2}(\mathbb{R})$, 
where $H_{n}(x)$ is the Hermite polynomial of variable $x$, 
$\gamma_{n}=\pi^{-1/4}(2^{n}n!)^{-1/2}$, and 
$w=\sqrt{m\omega/\hbar}$ 
for the frequency $\omega$ of a one-mode photon. 
We omit the tensor sign, $\otimes$, 
from the tensor product, $|s\rangle\otimes |n\rangle$ 
with $s=\uparrow, \downarrow$, $n=0, 1, 2, \cdots$, 
and use a compact notation, $|s\rangle|n\rangle$, 
for the tensor product. 
We respectively denote by $a$ and $a^{\dagger}$ 
the annihilation and creation operators of one-mode photon 
defined by $a|0\rangle:=0$, $a|n\rangle:=\sqrt{n}|n-1\rangle$, 
and $a^{\dagger}|n\rangle:=\sqrt{n+1}|n+1\rangle$. 
The spin-annihilation operator $\sigma_{-}$ and 
the spin-creation operator $\sigma_{+}$ are defined 
by $\sigma_{\pm}:=\frac{1}{2}\left(\sigma_{x}\pm i\sigma_{y}\right)$. 
The identity $2\times 2$ matrix $\sigma_{0}$ is given by 
$\sigma_{0}=\sigma_{+}\sigma_{-}+\sigma_{-}\sigma_{+}$.

\section{Some Reviews and Our Problem}
\label{sec:problem}

We introduce the parameters $\omega$ and $g$, 
respectively, playing roles of 
a frequency of a one-mode photon 
in a cavity and a coupling strength between an artificial 
two-level atom and the photon in the cavity. 
For every frequency $\omega$ and coupling strength $g$, 
the Hamiltonian of the generalized quantum Rabi model reads  
\begin{equation}
H_{\mbox{\tiny GQR}}(\omega,g)
:=H_{\mathrm{atm}}(\varepsilon)
+H_{\mathrm{ptn}}(\omega)
+\hbar g\sigma_{x}\left(a+a^{\dagger}\right) 
\label{eq:quantum-Rabi-Hamiltonian_0}
\end{equation}
with the two-level atom Hamiltonian $H_{\mathrm{atm}}(\varepsilon)$ 
and the one-mode photon Hamiltonian  $H_{\mathrm{ptn}}(\omega)$ 
defined by 
$$
H_{\mathrm{atm}}(\varepsilon)
:=\frac{\hbar}{2}\left(\omega_{\mathrm{a}}\sigma_{z}
-\varepsilon\sigma_{x}\right)\quad 
\mbox{and}\quad 
H_{\mathrm{ptn}}(\omega)
:=
\hbar\omega\left(a^{\dagger}a+\frac{1}{2}\right),
$$ 
where $\hbar\omega_{\mathrm{a}}$ and $\hbar\varepsilon$ 
are respectively the tunnel splitting 
and energy bias between the states, 
$|\!\!\uparrow\rangle$ and $|\!\!\downarrow\rangle$, 
of the two-level atom.

We recall the expression of the photon annihilation operator $a$ 
and creation operator $a^{\dagger}$ 
using the position operator $x$ 
and momentum operator $p$: 
\begin{equation}
a=\sqrt{\frac{\omega}{2\hbar}}\, x
+i\sqrt{\frac{1}{2\hbar\omega}}\, p  
\quad\textrm{and}\quad
a^{\dagger}=\sqrt{\frac{\omega}{2\hbar}}\, x
-i\sqrt{\frac{1}{2\hbar\omega}}\, p. 
\label{eq:expression_xp}
\end{equation}
Then, we have another expression of the photon Hamiltonian 
$H_{\mathrm{ptn}}(\omega)$ as 
\begin{equation}
H_{\mathrm{ptn}}(\omega)
=\frac{1}{2}p^{2}+\frac{\omega^{2}}{2}x^{2}
\label{eq:another-representation-H_ptn}
\end{equation}
using the canonical commutation relation 
$[x,p]=i\hbar$.

In the case where the energy bias is zero 
(i.e., $\varepsilon=0$), the generalized quantum Rabi 
Hamiltonian $H_{\mbox{\tiny GQR}}(\omega,g)$ 
becomes the quantum Rabi Hamiltonian. 
We denote it by $H_{\mbox{\tiny QR}}(\omega,g)$. 
The quantum Rabi Hamiltonian 
$H_{\mbox{\tiny QR}}(\omega,g)$ has 
the parity symmetry,
\begin{equation}
\left[ H_{\mbox{\tiny QR}}(\omega,g)\,\,\, ,\,\,\, 
\Pi\right]=0,
\label{eq:parity-symmetry}
\end{equation}
for the parity operator $\Pi=(-1)^{a^{\dagger}a}\sigma_{z}$. 

To introduce the form of the generalized quantum Rabi Hamiltonian 
$\mathcal{H}_{\mbox{\tiny GQR}}(\omega_{\mathrm{c}},\mathrm{g})$ 
that we consider in this paper, 
we define a unitary matrix $U_{xz}$ by 
\begin{equation}
U_{xz}:=\frac{1}{\sqrt{2}}
\left(\begin{array}{cc}
1 & 1 \\ 
-1 & 1
\end{array}\right).
\label{eq:unitary-matrix_U_xz}
\end{equation}
The Hamiltonian 
$\mathcal{H}_{\mbox{\tiny GQR}}(\omega,g)$ 
is given by  
\begin{eqnarray}
\mathcal{H}_{\mbox{\tiny GQR}}(\omega,g)
&:=& 
U_{xz}H_{\mbox{\tiny GQR}}(\omega,g)U_{xz}^{*}
\nonumber \\ 
&=&
\mathcal{H}_{\mathrm{atm}}(\varepsilon)
+H_{\mathrm{ptn}}(\omega)
+\hbar g\sigma_{z}\left( a+a^{\dagger}\right)
\label{eq:quantum-Rabi-Hamiltonian}
\end{eqnarray}
with the atom Hamiltonian $\mathcal{H}_{\mathrm{atm}}(\varepsilon)$ given by 
$$
\mathcal{H}_{\mathrm{atm}}(\varepsilon)
:=-\frac{\hbar}{2}\left(\omega_{\mathrm{a}}\sigma_{x}
+\varepsilon\sigma_{z}\right). 
$$ 
In this paper, we employ the one-mode photon frequency $\omega_{\mathrm{c}}$ 
and the coupling strength $\mathrm{g}$ 
as parameters $\omega$ and $g$, 
respectively, 
and we also call 
$\mathcal{H}_{\mbox{\tiny GQR}}(\omega_{\mathrm{c}},\mathrm{g})
:=U_{xz}H_{\mbox{\tiny GQR}}(\omega_{\mathrm{c}},\mathrm{g})U_{xz}^{*}$ 
the \textit{generalized quantum Rabi Hamiltonian}, 
and $\mathcal{H}_{\mbox{\tiny QR}}(\omega_{\mathrm{c}},\mathrm{g})
:=U_{xz}H_{\mbox{\tiny QR}}(\omega_{\mathrm{c}},\mathrm{g})U_{xz}^{*}$ 
the \textit{quantum Rabi Hamiltonian}.  
Using Eq.(\ref{eq:parity-symmetry}), 
we have the parity symmetry in the case $\varepsilon=0$,  
\begin{equation}
0=U_{xz}\left[ H_{\mbox{\tiny QR}}(\omega_{\mathrm{c}},\mathrm{g})\,\,\, ,\,\,\, 
\Pi\right]U_{xz}=
\left[\mathcal{H}_{\mbox{\tiny QR}}(\omega_{\mathrm{c}},\mathrm{g})\,\,\, ,\,\,\, 
\mathcal{P}\right],
\label{eq:parity-symmetry'}
\end{equation}
where $\mathcal{P}=-(-1)^{a^{\dagger}a}\sigma_{x}$. 
This determines the form of 
the eigenstates of the quantum Rabi Hamiltonian 
$\mathcal{H}_{\mbox{\tiny QR}}(\omega_{\mathrm{c}},\mathrm{g})$ 
as  
\begin{equation}
|\!\!\uparrow\rangle 
\biggl(|\mathrm{even}\rangle+|\mathrm{odd}\rangle\biggr)
\pm
|\!\!\downarrow\rangle
\biggl(|\mathrm{even}\rangle-|\mathrm{odd}\rangle\biggr)
\label{eq:form-eigenstates}
\end{equation}
for proper states $|\mathrm{even}\rangle$ 
and $|\mathrm{odd}\rangle$ with the individual forms,
$$
|\mathrm{even}\rangle=
\sum_{n:\mathrm{even}}c_{n}^{\mathrm{even}}|n\rangle\quad 
\mbox{and}\quad 
|\mathrm{odd}\rangle=
\sum_{n:\mathrm{odd}}c_{n}^{\mathrm{odd}}|n\rangle.
$$ 

Physically based on the argument in Ref.\cite{AN}, 
the Schr\"{o}dinger-cat-likeness appears 
in the \textit{adiabatic approximation} for the quantum Rabi Hamiltonian 
$\mathcal{H}_{\mbox{\tiny QR}}(\omega_{\mathrm{g}},\mathrm{g})$ 
(i.e., $\varepsilon=0$): all the eigenstates of 
the quantum Rabi Hamiltonian can be approximated by 
the Schr\"{o}dinger-cat-like states, 
\begin{equation}
\left\{
\begin{array}{ll}
{\displaystyle 
\frac{1}{\sqrt{2}}
\Biggl(|\!\!\uparrow\rangle 
D\left(-\mathrm{g}/\omega_{\mathrm{c}}\right)
|n\rangle
+|\!\!\downarrow\rangle 
D\left(\mathrm{g}/\omega_{\mathrm{c}}\right)
|n\rangle
\Biggr)}, \\ 
\qquad \\  
{\displaystyle \frac{1}{\sqrt{2}}
\Biggl(
|\!\!\uparrow\rangle 
D\left(-\mathrm{g}/\omega_{\mathrm{c}}\right)
|n\rangle
-|\!\!\downarrow\rangle 
D\left(\mathrm{g}/\omega_{\mathrm{c}}\right)
|n\rangle
\Biggr)},  
\end{array}
\right.
\label{eq:approximation_AN}
\end{equation} 
in the deep-strong coupling regime. 
Here, $D(\mathrm{g}/\omega_{\mathrm{c}})$ is the displacement operator 
defined by $D(\mathrm{g}/\omega_{\mathrm{c}})
:=\exp\left[\mathrm{g}\left(a^{\dagger}-a\right)/\omega_{\mathrm{c}}\right]$. 
Eqs.(\ref{eq:approximation_AN}) are well known 
as the adiabatic approximation (e.g., see Eq.(5) of Ref.\cite{irish}). 
The eigenenergies of the both eigenstates 
in Eq.(\ref{eq:approximation_AN}) are almost 
$\hbar\omega_{\mathrm{c}}(n+1/2)-\hbar\mathrm{g}^{2}/\omega_{\mathrm{c}}$; 
but, every true eigenstates are non-degenerate besides some cases. 
For instance, the adiabatically approximated eigenstates with the lowest energy 
$\hbar\omega_{\mathrm{c}}/2-\hbar\mathrm{g}^{2}/\omega_{\mathrm{c}}$ 
apparently seem to be degenerate; however, 
the ground state of the quantum Rabi Hamiltonian actually is unique 
for every coupling strength $\mathrm{g}$ \cite{HH}. 
  
As shown in Ref.\cite{HMS,hirokawa15}, 
the adiabatic approximation given by Eq.(\ref{eq:approximation_AN}) 
is mathematically justified in the following:  
We define the unitary operator 
$U(\mathrm{g}/\omega_{\mathrm{c}})$ by 
$U(\mathrm{g}/\omega_{\mathrm{c}})
:=\sigma_{+}\sigma_{-}D(\mathrm{g}/\omega_{\mathrm{c}})
+\sigma_{-}\sigma_{+}D(-\mathrm{g}/\omega_{\mathrm{c}})$. 
We note the equation, 
$U(\mathrm{g}/\omega_{\mathrm{c}})^{*}
=U(-\mathrm{g}/\omega_{\mathrm{c}})$. 
Then, we obtain the unitary transformation, 
\begin{eqnarray}
&{}& 
U(\mathrm{g}/\omega_{\mathrm{c}})
\left(\mathcal{H}_{\mbox{\tiny GQR}}(\omega_{\mathrm{c}},\mathrm{g})
+\hbar\mathrm{g}^{2}/\omega_{\mathrm{c}}\right)
U(\mathrm{g}/\omega_{\mathrm{c}})^{*} 
\nonumber \\ 
&=&H_{\mathrm{ptn}}(\omega_{\mathrm{c}})
-\frac{\hbar}{2}\varepsilon\sigma_{z}
-\frac{\hbar}{2}\omega_{\mathrm{a}}
\left\{
\sigma_{+}D\left(\mathrm{g}/\omega_{\mathrm{c}}\right)^{2} 
+\sigma_{-}D\left(-\mathrm{g}/\omega_{\mathrm{c}}\right)^{2}
\right\}, 
\label{eq:unitary-trans-GQR-Hamiltonian}
\end{eqnarray}
where $-\hbar\mathrm{g}^{2}/\omega_{\mathrm{c}}$ is the self-energy. 
Taking the strong coupling limit $\mathrm{g}\to\infty$, 
we have the limits, 
$\lim_{\mathrm{g}\to\infty}\hbar\mathrm{g}^{2}/\omega_{\mathrm{c}}=\infty$ 
and $\lim_{\mathrm{g}\to\infty}\mathrm{g}/\omega_{\mathrm{c}}=\infty$. 
Thus, the energy $\hbar\mathrm{g}^{2}/\omega_{\mathrm{c}}$ 
plays a role of a counter term for mass renormalization 
(i.e., for the bare-photon divergence) 
in the strong coupling limit. 
The displacement operators 
$D\left(\pm\mathrm{g}/\omega_{\mathrm{c}}\right)$ 
decay to the zero operator in a mathematically proper sense 
\cite{HMS,hirokawa15} 
in the strong-coupling limit. 
Developing this fact and using Theorem VIII.19(a) of Ref.\cite{RS1}, 
we can prove that the unitarily transformed Hamiltonian 
$U(\mathrm{g}/\omega_{\mathrm{c}})
\left(\mathcal{H}_{\mbox{\tiny GQR}}(\omega_{\mathrm{c}},\mathrm{g})
+\hbar\mathrm{g}^{2}/\omega_{\mathrm{c}}
\right)U(\mathrm{g}/\omega_{\mathrm{c}})^{*}$ 
converges to the Hamiltonian 
$H_{\mathrm{ptn}}(\omega_{\mathrm{c}})-\hbar\varepsilon\sigma_{z}/2$
in the norm resolvent sense:
$$
U(\mathrm{g}/\omega_{\mathrm{c}})
\left(\mathcal{H}_{\mbox{\tiny GQR}}(\omega_{\mathrm{c}},\mathrm{g})
+\hbar\mathrm{g}^{2}/\omega_{\mathrm{c}}
\right)U(\mathrm{g}/\omega_{\mathrm{c}})^{*} 
\stackrel{\mathrm{n.r.s.}\,\,}{{-\!\!-\!\!\!}\longrightarrow} 
H_{\mathrm{ptn}}(\omega_{\mathrm{c}})-\frac{\hbar}{2}\varepsilon\sigma_{z}
\,\,\, 
\mbox{as $\mathrm{g}\to\infty$}. 
$$
Thanks to Theorem VIII.23(b) of Ref.\cite{RS1}, 
each eigenstate of 
the Hamiltonian $\mathcal{H}_{\mbox{\tiny QR}}(\omega_{\mathrm{c}},\mathrm{g})$ 
is well approximated by that of the Hamiltonian 
$U(\mathrm{g}/\omega_{\mathrm{c}})^{*}
(H_{\mathrm{ptn}}(\omega_{\mathrm{c}})
-\hbar\varepsilon\sigma_{z}/2)
U(\mathrm{g}/\omega_{\mathrm{c}})
-\hbar\mathrm{g}^{2}/\omega_{\mathrm{c}}$. 
This mathematical procedure with $\varepsilon=0$ secures 
the adiabatic-approximation formulas given by Eq.(\ref{eq:approximation_AN}). 
On the other hand, in the case where $\varepsilon\ne 0$, 
all the eigenstates of the generalized quantum Rabi Hamiltonian 
$\mathcal{H}_{\mbox{\tiny GQR}}(\omega_{\mathrm{c}},\mathrm{g})$ 
is well approximated by the states, 
\begin{equation}
\left\{
\begin{array}{ll}
{\displaystyle 
|\!\!\uparrow\rangle 
D\left(-\mathrm{g}/\omega_{\mathrm{c}}\right)
|n\rangle}, \\ 
\qquad \\  
{\displaystyle 
|\!\!\downarrow\rangle 
D\left(\mathrm{g}/\omega_{\mathrm{c}}\right)
|n\rangle
}.
\end{array}
\right.
\label{eq:approximation_0}
\end{equation}
The first adiabatically approximated eigenstates 
$|\!\!\uparrow\rangle 
D\left(-\mathrm{g}/\omega_{\mathrm{c}}\right)
|n\rangle$ 
gives the eigenenergy 
$\hbar\omega_{\mathrm{c}}\left(n+1/2\right)-\hbar\varepsilon/2
-\hbar\mathrm{g}^{2}/\omega_{\mathrm{c}}$, 
and the second one gives 
the eigenenergy 
$\hbar\omega_{\mathrm{c}}\left(n+1/2\right)+\hbar\varepsilon/2
-\hbar\mathrm{g}^{2}/\omega_{\mathrm{c}}$. 
Following the adiabatic-approximation formulas (\ref{eq:approximation_0}), 
whether the energy bias is positive or negative causes 
an energy level crossing.  
At last, we realized that i) the limit Hamiltonian, 
$U(\mathrm{g}/\omega_{\mathrm{c}})^{*}
(H_{\mathrm{ptn}}(\omega_{\mathrm{c}})-\hbar\varepsilon\sigma_{z}/2
)U(\mathrm{g}/\omega_{\mathrm{c}})
-\hbar\mathrm{g}^{2}/\omega_{\mathrm{c}}$, 
as well as its eigenstates and eigenenergies 
does not include the tunnel splitting $\hbar\omega_{\mathrm{a}}$ 
of two-level atom, 
but the energy bias $\hbar\varepsilon$; 
ii) the eigenstates in Eq.(\ref{eq:approximation_AN}) 
are the Schr\"{o}dinger-cat-like, 
but the eigenstates in Eq.(\ref{eq:approximation_0}) are not. 

Here, we make a remark on a physical role 
of the displacement operator 
$D(\pm\mathrm{g}/\omega_{\mathrm{c}})$ 
to introduce our problem. 
The appearance of the displacement operator 
in Eqs.(\ref{eq:approximation_AN}) 
and (\ref{eq:approximation_0}) makes coherent states. 
However, they are for bare photons; and in fact, 
the photon-field fluctuation $\Delta\Phi$ increases 
the ground-state expectation 
$N_{0}^{\mbox{\tiny GQR}}(\omega_{\mathrm{c}},\mathrm{g})$  
of the number of photons. 
More precisely, Eqs.(\ref{eq:approximation_AN}) 
and (\ref{eq:approximation_0}) 
say that the ground-state expectation 
$N_{0}^{\mbox{\tiny GQR}}(\omega_{\mathrm{c}},\mathrm{g})
=\langle E_{0}^{\mbox{\tiny GQR}}|a^{\dagger}a
|E_{0}^{\mbox{\tiny GQR}}\rangle$ increases 
as the coupling strength $\mathrm{g}$ grows larger, 
i.e., $N_{0}^{\mbox{\tiny GQR}}(\omega_{\mathrm{c}},\mathrm{g})
\sim\mathrm{g}^{2}/\omega_{\mathrm{c}}^{2}$ as $\mathrm{g}\to\infty$, 
where $|E_{0}^{\mbox{\tiny GQR}}\rangle$ is the ground state 
of the generalized quantum Rabi Hamiltonian 
$\mathcal{H}_{\mbox{\tiny GQR}}(\omega_{\mathrm{c}},\mathrm{g})$. 
This increase results from the mathematical establishment of 
the adiabatic approximation. 
Actually, it is pushed up by the fluctuation $\Delta\Phi$ 
of the photon field $\Phi:=
(a+a^{\dagger})/\sqrt{2\omega_{\mathrm{c}}}$ 
in the ground state 
since the inequality,  
$$
(\Delta\Phi)^{2}
\le \frac{
2N_{0}^{\mbox{\tiny GQR}}(\omega_{\mathrm{c}},\mathrm{g})+1
}{\omega_{\mathrm{c}}},
$$ 
is obtained in the same way as in Appendix B of Ref.\cite{HMS}. 
The mathematical establishment of the adiabatic approximation 
also says that $(\Delta\Phi)^{2}\sim 
(1+4\mathrm{g}^{2}/\omega_{\mathrm{c}})/2\omega_{\mathrm{c}}$ 
as $\mathrm{g}\to\infty$ for $\varepsilon=0$; 
$(\Delta\Phi)^{2}\sim 
1/2\omega_{\mathrm{c}}$ 
as $\mathrm{g}\to\infty$ for $\varepsilon\ne 0$. 
Therefore, the Schr\"{o}dinger-cat-likeness is caused by 
bare photons, and it is not observable directly. 

We now try to derive physical states from the 
adiabatically approximated bare states given in 
Eqs.(\ref{eq:approximation_AN}) 
and (\ref{eq:approximation_0}). 
The Hamiltonians $H_{\pm\mbox{\tiny vH}}$ of the 
van Hove model for the neutral scalar field theory 
with a fixed sources \cite{van-hove} are given by 
$H_{\pm\mathrm{vH}}:=H_{\mathrm{ptn}}(\omega_{\mathrm{c}})
\pm\hbar\mathrm{g}(a+a^{\dagger})$. 
We denote by $|n_{\pm\mbox{\tiny vH}}\rangle$ the eigenstate of 
the van Hove Hamiltonians $H_{\pm\mathrm{vH}}$. 
Since each eigenstate is given by 
$|n_{\pm\mbox{\tiny vH}}\rangle
=D(\mp\mathrm{g}/\omega_{\mathrm{c}})|n\rangle$, 
the ground-state expectation is calculated as 
$\langle 0_{\pm\mbox{\tiny vH}}|a^{\dagger}a|0_{\pm\mbox{\tiny vH}}\rangle
=\mathrm{g}^{2}/\omega_{\mathrm{c}}^{2}$. 
It increases in association with the growth 
of the coupling strength $\mathrm{g}$ 
as it looks as it appears to be. 
However, we find unitary operators $U_{\pm\mbox{\tiny vH}}$ 
to derive physical states from bare states 
for the van Hove Hamiltonians $H_{\pm\mathrm{vH}}$, 
and then, 
we have the renormalized van Hove Hamiltonian given by 
$U_{\pm\mbox{\tiny vH}}^{*}
(H_{\pm\mbox{\tiny vH}}+\hbar\mathrm{g}^{2}/\omega_{\mathrm{c}})
U_{\pm\mbox{\tiny vH}}^{\mathrm{ren}}$. 
Here, the energy $-\hbar\mathrm{g}^{2}/\omega_{\mathrm{c}}$ is 
the self-energy of the van Hove Hamiltonian, 
and we have to make the so-called mass renormalization \cite{HT}. 
In addition to this, Ref.\cite{HT} tells us that 
the unitary operators are given by $U_{\pm\mbox{\tiny vH}}
=D(\mp\mathrm{g}/\omega_{\mathrm{c}})$, 
and each physical state $|n_{\pm\mbox{\tiny vH}}^{\mathrm{ren}}\rangle$ 
of the bare state $|n_{\pm\mbox{\tiny vH}}\rangle$ 
is given by $|n_{\pm\mbox{\tiny vH}}^{\mathrm{ren}}\rangle
=U_{\pm\mbox{\tiny vH}}^{*}|n_{\pm\mbox{\tiny vH}}\rangle$. 
Eventually, the physical state $|n_{\pm\mbox{\tiny vH}}^{\mathrm{ren}}\rangle$ 
gets itself satisfying 
$|n_{\pm\mbox{\tiny vH}}^{\mathrm{ren}}\rangle=|n\rangle$. 
The photon in the physical state 
$|n_{\pm\mbox{\tiny vH}}^{\mathrm{ren}}\rangle$ 
is the so-called dressed photon, 
which sometimes called real photon. 
Thus, we can expect no dressed photon 
in the physical ground state 
$|0_{\pm\mbox{\tiny vH}}^{\mathrm{ren}}\rangle$, i.e.,   
$\langle 0_{\pm\mbox{\tiny vH}}^{\mathrm{ren}}|
a^{\dagger}a
|0_{\pm\mbox{\tiny vH}}^{\mathrm{ren}}\rangle=0$. 
Therefore, $\langle 0_{\pm\mbox{\tiny vH}}|a^{\dagger}a|0_{\pm\mbox{\tiny vH}}\rangle
=\mathrm{g}^{2}/\omega_{\mathrm{c}}^{2}$ 
is the expectation value of the number of 
the bare photons including virtual photons 
in the bare ground sate. 

Following the argument in Complement B${}_{\mbox{\tiny III}}$.2 
of Ref.\cite{cohen-tannoudji}, 
we can think that the photons in the ground state 
are virtual photons since the ground state expectation 
$N_{0}^{\mbox{\tiny GQR}}(\omega_{\mathrm{c}},\mathrm{g})$ 
increases as the coupling strength $\mathrm{g}$ grows larger: 
$N_{0}^{\mbox{\tiny GQR}}(\omega_{\mathrm{c}},\mathrm{g})
\sim 
\mathrm{g}^{2}/\omega_{\mathrm{c}}^{2}$ 
as $\mathrm{g}\to\infty$.  
Using the representation in Ref.\cite{CS}, we define 
the annihilation operator $\alpha$ 
and creation operator $\alpha^{\dagger}$ by 
$\alpha^{\sharp}:=\sigma_{z}a^{\sharp}$. 
Then, we have the matrix-valued CCR, $[\alpha , \alpha^{\dagger}]=1$, 
and the expression, 
$$
\mathcal{H}_{\mbox{\tiny GQR}}(\omega_{\mathrm{c}},\mathrm{g})
=\hbar\omega_{\mathrm{c}}\left(\alpha^{\dagger}\alpha+\frac{1}{2}\right)
+\hbar\mathrm{g}\left(\alpha+\alpha^{\dagger}\right)
+\mathcal{H}_{\mathrm{atm}}(\varepsilon).
$$ 
In the sufficiently strong coupling regime, 
the atom Hamiltonian $\mathcal{H}_{\mathrm{atm}}(\varepsilon)$ 
can be regarded as the perturbation 
of the van Hove Hamiltonian 
$\hbar\omega_{\mathrm{c}}\left(\alpha^{\dagger}\alpha+1/2\right)
+\hbar\mathrm{g}\left(\alpha+\alpha^{\dagger}\right)$. 
This is the very idea of the adiabatic approximation.  
Based on this fact, 
in a similar way to the van Hove model's case, 
we define the unitary operator $U_{\mbox{\tiny GQR}}$ 
for deriving physical states from bare states by
$U_{\mbox{\tiny GQR}}:=\sigma_{+}\sigma_{-}U_{+\mbox{\tiny vH}}
+\sigma_{-}\sigma_{+}U_{-\mbox{\tiny vH}}$. 
Then, for each bare state $\psi$ of the 
generalized quantum Rabi Hamiltonian 
$\mathcal{H}_{\mbox{\tiny GQR}}(\omega_{\mathrm{c}},\mathrm{g})$, 
we have the physical eigenstates $\psi^{\mathrm{ren}}$ 
by $\psi^{\mathrm{ren}}=U_{\mbox{\tiny GQR}}^{*}\psi$. 
Since the renormalized Hamiltonian 
$\mathcal{H}_{\mbox{\tiny GQR}}^{\mathrm{ren}}(\omega_{\mathrm{c}},\mathrm{g})$ 
of the generalized quantum Rabi Hamiltonian 
$\mathcal{H}_{\mbox{\tiny GQR}}(\omega_{\mathrm{c}},\mathrm{g})$ 
is given by 
$\mathcal{H}_{\mbox{\tiny GQR}}^{\mathrm{ren}}(\omega_{\mathrm{c}},\mathrm{g})
=U_{\mbox{\tiny GQR}}^{*}
\left(
\mathcal{H}_{\mbox{\tiny GQR}}(\omega_{\mathrm{c}},\mathrm{g})
+\hbar\mathrm{g}^{2}/\omega_{\mathrm{c}}
\right)
U_{\mbox{\tiny GQR}}$ 
and $U_{\mbox{\tiny GQR}}=U(\mathrm{g}/\omega_{\mathrm{c}})$, 
we have  
\begin{equation}
\mathcal{H}_{\mbox{\tiny GQR}}^{\mathrm{ren}}(\omega_{\mathrm{c}},\mathrm{g})
= 
H_{\mathrm{ptn}}(\omega_{\mathrm{c}})
-\frac{\hbar}{2}\varepsilon\sigma_{z}
-\frac{\hbar}{2}\omega_{\mathrm{a}}
\left\{
\sigma_{+}U_{+\mbox{\tiny vH}}^{*}U_{-\mbox{\tiny vH}}
+\sigma_{-}U_{-\mbox{\tiny vH}}^{*}U_{+\mbox{\tiny vH}}
\right\}. 
\label{eq:renormalized-GQR-Hamiltonian}
\end{equation} 
We immediately realize that RHS of Eq.(\ref{eq:renormalized-GQR-Hamiltonian}) 
is RHS of Eq.(\ref{eq:unitary-trans-GQR-Hamiltonian}). 
In the case $\varepsilon=0$, 
using Eq.(\ref{eq:approximation_AN}), 
all the normalized eigenstates $\psi$ of the quantum Rabi Hamiltonian 
$\mathcal{H}_{\mbox{\tiny QR}}(\omega_{\mathrm{c}},\mathrm{g})$ 
are approximated by $(|\!\!\uparrow\rangle|n_{+\mbox{\tiny vH}}\rangle
+|\!\!\downarrow\rangle|n_{-\mbox{\tiny vH}}\rangle)/\sqrt{2}$ 
or $(|\!\!\uparrow\rangle|n_{+\mbox{\tiny vH}}\rangle
-|\!\!\downarrow\rangle|n_{-\mbox{\tiny vH}}\rangle)/\sqrt{2}$, 
$n=0, 1, 2, \cdots$. 
Thus, the physical eigenstates $\psi^{\mathrm{ren}}$ (i.e., 
eigenstates of the renormalized Hamiltonian 
$\mathcal{H}_{\mbox{\tiny QR}}^{\mathrm{ren}}(\omega_{\mathrm{c}},\mathrm{g})$ 
are approximated by 
\begin{equation}
\left\{
\begin{array}{ll}
{\displaystyle 
\frac{1}{\sqrt{2}}
\Biggl(|\!\!\uparrow\rangle 
|n_{+\mbox{\tiny vH}}^{\mathrm{ren}}\rangle
+|\!\!\downarrow\rangle 
|n_{-\mbox{\tiny vH}}^{\mathrm{ren}}\rangle
\Biggr)
=
\frac{1}{\sqrt{2}}
\Biggl(|\!\!\uparrow\rangle 
+|\!\!\downarrow\rangle 
\Biggr)
|n\rangle
}, \\ 
\qquad \\  
{\displaystyle \frac{1}{\sqrt{2}}
\Biggl(
|\!\!\uparrow\rangle 
|n_{+\mbox{\tiny vH}}^{\mathrm{ren}}\rangle
-|\!\!\downarrow\rangle 
|n_{-\mbox{\tiny vH}}^{\mathrm{ren}}\rangle
\Biggr)
=
\frac{1}{\sqrt{2}}
\Biggl(
|\!\!\uparrow\rangle 
-|\!\!\downarrow\rangle 
\Biggr)
|n\rangle
}. 
\end{array}
\right.
\label{eq:approximation_AN'}
\end{equation}
In the same way, in the case $\varepsilon\ne 0$, 
using Eq.(\ref{eq:approximation_0}), 
we have the adiabatically approximated physical eigenstates 
\begin{equation}
\left\{
\begin{array}{ll}
{\displaystyle 
|\!\!\uparrow\rangle 
|n_{+\mbox{\tiny vH}}^{\mathrm{ren}}\rangle
=|\!\!\uparrow\rangle 
|n\rangle}, \\ 
\qquad \\  
{\displaystyle 
|\!\!\downarrow\rangle 
|n_{-\mbox{\tiny vH}}^{\mathrm{ren}}\rangle
=|\!\!\downarrow\rangle 
|n\rangle
},
\end{array}
\right.
\label{eq:approximation_0'} 
\end{equation}
for sufficiently large coupling strength. 
Thus, the adiabatically approximated physical states in 
Eqs.(\ref{eq:approximation_AN'}) 
and (\ref{eq:approximation_0'}) can no longer make 
any coherent sate of dressed photons, 
and are no longer macroscopic. 
That is, Schr\"{o}dinger-cat-likeness does not appear 
in those physical eigenstates.

As observed above, if we employ the unitary operator 
$U_{\mbox{\tiny GQR}}$ to derive physical states 
from bare states for the (generalized) quantum Rabi model, 
physical eigenstates are approximated by 
eigenstates of the free part of the (generalized) quantum Rabi 
Hamiltonian (i.e., $\mathrm{g}=0$). 
Moreover, the adiabatic approximations given by Eqs.(\ref{eq:approximation_AN'}) 
and (\ref{eq:approximation_0'}) tell us that 
the (generalized) quantum Rabi Hamiltonian cannot 
make us expect any dressed photon in the physical ground state 
even for sufficiently large coupling strength. 
Here, we point out the following properties: 
\begin{description}
\item[P1)] The derivations of Eqs.(\ref{eq:approximation_AN}) and 
(\ref{eq:approximation_0}) 
tell us that 
whether the adiabatically approximated eigenstates of the generalized quantum 
Rabi Hamiltonian are formed as 
the Schr\"{o}dinger-cat-like entangled states or not 
depends on whether the energy-bias parameter is zero or non-zero, 
in other words, whether the parity symmetry given by Eq.(\ref{eq:parity-symmetry'}) 
is conserved or not. 
\item[P2)] The limit Hamiltonian in the norm resolvent sense says that 
the effect of the tunnel splitting of the two-level atom can be ignored 
in the adiabatic approximations. 
\item[P3)] Following the theory of van Hove model, 
the Schr\"{o}dinger-cat-likeness disappears from 
the adiabatically approximated physical eigenstates. 
\end{description}
We are interested in a vestige of the Schr\"{o}dinger-cat-likeness 
of bare photons in the physical eigenstates. 

Meanwhile, we have to consider the quadratic interaction (i.e., $A^{2}$-term) 
of the photon field in the case where our physical system 
of the two-level atom interacting with the one-mode photon field 
is in the very strong coupling regime such as the deep-strong coupling regime. 
For such a situation, we should consider the Hamiltonian 
\begin{equation}
\mathcal{H}_{A^{2}}(\varepsilon):=
\mathcal{H}_{\mbox{\tiny GQR}}(\omega_{\mathrm{c}},\mathrm{g})
+\hbar\mathrm{g}C_{\mathrm{g}}
\left(a+a^{\dagger}\right)^{2}, 
\label{eq:total-Hamiltonian}
\end{equation}
where $C_{\mathrm{g}}$ is a positive function of 
the coupling strength $\mathrm{g}$ satisfying 
$\lim_{\mathrm{g}\to\infty}\mathrm{g}C_{\mathrm{g}}=\infty$. 
We assume the following conditions:
\begin{equation}
\lim_{\mathrm{g}\to\infty}\mathrm{g}^{-1/3}C_{\mathrm{g}}=\infty,
\label{eq:assumption}
\end{equation}
and there is a non-negative constant $C_{\infty}$ so that 
\begin{equation}
\lim_{\mathrm{g}\to\infty}\mathrm{g}^{-1}C_{\mathrm{g}}=C_{\infty}.
\label{eq:assumption'}
\end{equation}
For instance, if we estimate 
$C_{\mathrm{g}}$ at $(\mathrm{const})\times\mathrm{g}$, 
then Eqs.(\ref{eq:assumption}) and (\ref{eq:assumption'}) hold. 

In this paper, we investigate an effect caused by the $A^{2}$-term 
in the physical eigenstates 
of our total Hamiltonian $\mathcal{H}_{A^{2}}(\varepsilon)$. 
To obtain the physical states, 
we employ the (meson) pair theory in nuclear physics 
\cite{HT,wentzel1,wentzel2,BR,KM}. 
Then, we have a unitary operator $U_{A^{2}}$ such that 
the unitarily transformed Hamiltonian 
$U_{A^{2}}^{*}\mathcal{H}_{A^{2}}(\varepsilon)U_{A^{2}}$ 
becomes the renormalized Hamiltonian for the physical 
eigenstates. 
Following (meson) pair theory, 
we obtain the unitary operator $U_{A^{2}}$ given by 
the Hopfield-Bogoliubov transformation $U_{\mbox{\tiny HB}}$, 
i.e., $U_{A^{2}}=U_{\mbox{\tiny HB}}$,  
as shown in Ref.\cite{HMS}, 
so that we obtain the renormalized Hamiltonian 
$\mathcal{H}_{A^{2}}^{\mathrm{ren}}(\varepsilon)$ as 
\begin{equation}
\mathcal{H}_{A^{2}}^{\mathrm{ren}}(\varepsilon)
:=
U_{A^{2}}^{*}\mathcal{H}_{A^{2}}(\varepsilon)U_{A^{2}}
=\mathcal{H}_{\mathrm{atm}}(\varepsilon)
+H_{\mathrm{ptn}}(\omega_{\mathrm{g}})
+\hbar\widetilde{\mathrm{g}}\sigma_{z}\left(a+a^{\dagger}\right) 
=
\mathcal{H}_{\mbox{\tiny GQR}}(\omega_{\mathrm{g}},\widetilde{\mathrm{g}}), 
\qquad 
\label{eq:renormalized-Hamiltonian}
\end{equation}
where $\omega_{\mathrm{g}}$ and 
$\widetilde{\mathrm{g}}$ 
are respectively renormalized photon frequency 
and the renormalized coupling strength 
given by 
$$
\omega_{\mathrm{g}}=
\sqrt{\omega_{\mathrm{c}}^{2}+4\omega_{\mathrm{c}}\mathrm{g}C_{\mathrm{g}}}
\quad \mbox{and}\quad 
\widetilde{\mathrm{g}}
=\mathrm{g}\sqrt{\frac{\omega_{\mathrm{c}}}{\omega_{\mathrm{g}}}}.
$$ 
We briefly review how to obtain 
Eq.(\ref{eq:renormalized-Hamiltonian}) 
in the next section. 

Similarly to the argument for the generalized quantum Rabi model, 
we introduce the unitary operator 
$U(\widetilde{\mathrm{g}}/\omega_{\mathrm{g}})$ by 
\begin{equation}
U(\widetilde{\mathrm{g}}/\omega_{\mathrm{g}})
:=\sigma_{+}\sigma_{-}D(\widetilde{\mathrm{g}}/\omega_{\mathrm{g}})
+\sigma_{-}\sigma_{+}D(-\widetilde{\mathrm{g}}/\omega_{\mathrm{g}}),
\label{eq:unitary-operator-van_Hove}
\end{equation} 
where $D(\widetilde{\mathrm{g}}/\omega_{\mathrm{g}})$ 
is the displacement operator 
defined by $D(\widetilde{\mathrm{g}}/\omega_{\mathrm{g}})
:=\exp\left[\widetilde{\mathrm{g}}
\left(a^{\dagger}-a\right)/\omega_{\mathrm{g}}\right]$. 
Then, we have the unitary transformation, 
\begin{eqnarray}
&{}& 
U\left(\widetilde{\mathrm{g}}/\omega_{\mathrm{g}}\right)
\left(\mathcal{H}_{A^{2}}^{\mathrm{ren}}(\varepsilon)
+\hbar\widetilde{\mathrm{g}}^{2}/\omega_{\mathrm{g}}\right)
U\left(\widetilde{\mathrm{g}}/\omega_{\mathrm{g}}\right)^{*} 
\nonumber \\ 
&=&
H_{\mathrm{ptn}}(\omega_{\mathrm{g}})
-\frac{\hbar}{2}\varepsilon\sigma_{z}
-\frac{\hbar}{2}\omega_{\mathrm{a}}
\left\{
\sigma_{+}D\left(\widetilde{\mathrm{g}}/\omega_{\mathrm{g}}\right)^{2} 
+\sigma_{-}D\left(-\widetilde{\mathrm{g}}/\omega_{\mathrm{g}}\right)^{2}
\right\}.
\label{eq:unitary-trans}
\end{eqnarray} 
For Eq.(\ref{eq:unitary-trans}), we realize the followings: 
Since the limit 
\begin{equation}
\lim_{\mathrm{g}\to\infty}
\hbar\frac{\widetilde{\mathrm{g}}^{2}}{\omega_{\mathrm{g}}}
=\frac{\hbar}{4C_{\infty}}
\label{eq:limit_of_self-energy}
\end{equation} 
follows from Eq.(\ref{eq:assumption'}), 
the self-energy $\hbar\widetilde{\mathrm{g}}^{2}/\omega_{\mathrm{g}}$ 
does not work as a counter term 
when we take the strong coupling limit $\mathrm{g}\to\infty$. 
Instead, we meet a trouble of divergence for 
the Hamiltonian $H_{\mathrm{ptn}}(\omega_{\mathrm{g}})$ 
due to the divergence $\lim_{\mathrm{g}\to\infty}\omega_{\mathrm{g}}
=\infty$. 
In addition to this trouble, 
the third term of RHS of Eq.(\ref{eq:unitary-trans}) 
does not vanish as the coupling strength $\mathrm{g}$ 
tends to the infinity 
because the displacement operators 
$D(\widetilde{\mathrm{g}}/\omega_{\mathrm{g}})$ 
and 
$D(\widetilde{\mathrm{g}}/\omega_{\mathrm{g}})^{*}$ 
do not decay to zero because of the limit, 
$$
\lim_{\mathrm{g}\to\infty}
\frac{\widetilde{\mathrm{g}}}{\,\,\omega_{\mathrm{g}}}
=\lim_{\mathrm{g}\to\infty}
\omega_{\mathrm{c}}^{-1/4}
\left(
\omega_{\mathrm{c}}\mathrm{g}^{-4/3}
+4\mathrm{g}^{-1/3}C_{\mathrm{g}}
\right)^{-3/4}=0,
$$  
by Eq.(\ref{eq:assumption}). 

In this paper, coping with the trouble and difference, 
we will consider the properties corresponding to 
\textbf{P1}, \textbf{P2}, and \textbf{P3} 
for our total Hamiltonian $\mathcal{H}^{\mathrm{ren}}_{A^{2}}(\varepsilon)$.

\section{From Bare Eigenstates to Physical Eigenstates} 

Following (meson) pair theory \cite{HT,wentzel1,wentzel2,BR,KM}, 
we obtain physical eigenstates from bare ones. 
We review it in brief. 
For more details on how to apply (meson) pair theory to our model, 
see the argument in Ref.\cite{HMS}. 
Our Hamiltonian $\mathcal{H}_{A^{2}}(\varepsilon)$ 
has the matrix representation as 
$$
\mathcal{H}_{A^{2}}(\varepsilon)
=
\left(
\begin{array}{cccc}
H_{A^{2}}^{+}-\hbar\varepsilon/2 & -\hbar\omega_{\mathrm{a}}/2 \\ 
-\hbar\omega_{\mathrm{a}}/2 & H_{A^{2}}^{-}+\hbar\varepsilon/2
\end{array}
\right), 
$$
where 
$$
H_{A^{2}}^{\pm}=
H_{\mathrm{ptn}}(\omega_{\mathrm{c}})
\pm\hbar\mathrm{g}\left( a+ a^{\dagger}\right) 
+\hbar C_{\mathrm{g}}\mathrm{g}
\left( a+ a^{\dagger}\right)^{2}. 
$$
Using Eqs.(\ref{eq:expression_xp}) and 
(\ref{eq:another-representation-H_ptn}), we can rewrite 
the Hamiltonians $H_{A^{2}}^{\pm}$ as 
$$
H_{A^{2}}^{\pm}=
\frac{1}{2}p^{2}+\frac{\omega_{\mathrm{c}}^{2}}{2}x^{2}
\pm\hbar\mathrm{g}\sqrt{\frac{2\omega_{\mathrm{c}}}{\hbar}}\, x
+2C_{\mathrm{g}}\mathrm{g}\omega_{\mathrm{c}}x^{2} 
= 
\frac{1}{2}p^{2}+\frac{\omega_{\mathrm{g}}^{2}}{2}x^{2}
\pm\hbar\widetilde{\mathrm{g}}\sqrt{\frac{2\omega_{\mathrm{g}}}{\hbar}}\, x. 
$$ 
We define new photon's annihilation operator $b$ 
and creation operator $b^{\dagger}$ by
\begin{equation}
b:=\sqrt{\frac{\omega_{\mathrm{g}}}{2\hbar}}\, x
+i\sqrt{\frac{1}{2\hbar\omega_{\mathrm{g}}}}\, p  
\quad\textrm{and}\quad
b^{\dagger}:=\sqrt{\frac{\omega_{\mathrm{g}}}{2\hbar}}\, x
-i\sqrt{\frac{1}{2\hbar\omega_{\mathrm{g}}}}\, p. 
\label{eq:expression_xp'}
\end{equation}
Then, we have expression of the Hamiltonians $H_{A^{2}}^{\pm}$ 
using the new annihilation and creation operators, 
$b$ and $b^{\dagger}$, as  
\begin{equation}
H_{A^{2}}^{\pm}=
\hbar\omega_{\mathrm{g}}\left(b^{\dagger}b+\frac{1}{2}\right)
\pm\hbar\widetilde{\mathrm{g}}\left(b+b^{\dagger}\right).
\label{eq:key-representation}
\end{equation}
Making the correspondence between the normalized eigenstates of 
the Hamiltonian 
$(1/2)p^{2}+(\omega_{\mathrm{c}}^{2}/2)x^{2}$ 
to those of the Hamiltonian 
$(1/2)p^{2}+(\omega_{\mathrm{g}}^{2}/2)x^{2}$, 
we eventually obtain the so-called Hopfield-Bogoliubov transformation 
$U_{\mbox{\tiny HB}}$, 
and then, reach the unitary transformation, 
$$
\begin{cases}
U_{\mbox{\tiny HB}}aU_{\mbox{\tiny HB}}^{*}=b
=\frac{1}{2}(c_{1}+c_{2})a+\frac{1}{2}(c_{1}-c_{2})a^{\dagger}, \\ 
U_{\mbox{\tiny HB}}a^{\dagger}U_{\mbox{\tiny HB}}^{*}=b^{\dagger}
=\frac{1}{2}(c_{1}-c_{2})a+\frac{1}{2}(c_{1}+c_{2})a^{\dagger}, 
\end{cases}
$$
where 
$c_{1}=\sqrt{\omega_{\mathrm{g}}/\omega_{\mathrm{c}}}$ 
and $c_{2}=\sqrt{\omega_{\mathrm{c}}/\omega_{\mathrm{g}}}$. 
We note that the Hopfield-Bogoliubov transformation $U_{\mbox{\tiny HB}}$ 
is concretely defined by Eqs.(50) and (57) of Ref.\cite {HMS} or 
Eqs.(12.17)-(12.19) of Ref.\cite{HT} 
in (meson) pair theory. 
Use the Hopfield-Bogoliubov transformation $U_{\mbox{\tiny HB}}$, 
and we obtain the unitary transformation
\begin{equation}
U_{\mbox{\tiny HB}}^{*}\mathcal{H}_{A^{2}}(\varepsilon)U_{\mbox{\tiny HB}}
=\mathcal{H}_{\mbox{\tiny GQR}}(\omega_{\mathrm{g}},\widetilde{\mathrm{g}}).
\label{eq:Hamiltonian-after-HB}
\end{equation}
This is our renormalized Hamiltonian 
$\mathcal{H}_{A^{2}}^{\mathrm{ren}}(\varepsilon)$ 
in Eq.(\ref{eq:renormalized-Hamiltonian}). 

We denote by $|\mathcal{E}_{\nu}(\varepsilon)\rangle$ eigenstates 
of our total Hamiltonian  $\mathcal{H}_{A^{2}}(\varepsilon)$ 
with eigenenergy $E_{\nu}$, i.e., 
$\mathcal{H}_{A^{2}}(\varepsilon)|\mathcal{E}_{\nu}(\varepsilon)\rangle
=E_{\nu}|\mathcal{E}_{\nu}(\varepsilon)\rangle$. 
We make the order of the eigenenergies as 
$E_{1}\le E_{2}\le \cdots\le 
E_{\nu}\le E_{\nu+1}\le \cdots$. 
These eigenstates, $|\mathcal{E}_{\nu}(\varepsilon)\rangle$, 
$\nu=0, 1, 2, \cdots$, are bare states. 
According to (meson) pair theory, 
we should derive the physical eigenstates 
$|\mathcal{E}_{\nu}^{\mathrm{ren}}(\varepsilon)\rangle$ form the bare ones 
by 
\begin{equation}
|\mathcal{E}_{\nu}^{\mathrm{ren}}(\varepsilon)\rangle
:=U_{\mbox{\tiny HB}}^{*}|\mathcal{E}_{\nu}(\varepsilon)\rangle. 
\label{eq:physical-eigenstaes}
\end{equation}
Then, we have 
$\mathcal{H}_{A^{2}}^{\mathrm{ren}}(\varepsilon)
|\mathcal{E}_{\nu}^{\mathrm{ren}}(\varepsilon)\rangle
=E_{\nu}|\mathcal{E}_{\nu}^{\mathrm{ren}}(\varepsilon)\rangle$.
In next section, we will give the adiabatic approximation 
for these physical eigenstates 
$|\mathcal{E}_{\nu}^{\mathrm{ren}}(\varepsilon)\rangle$ 
by taking the strong coupling limit.

\section{Schr\"{o}dinger-Cat-Likeness in Adiabatic Approximation}

In this section, we show the Schr\"{o}dinger-cat-likeness 
in the adiabatic approximation. 
We will give a mathematical proof of the adiabatic approximation 
in the next section.  

To make an energy renormalization for $H_{\mathrm{ptn}}(\omega_{\mathrm{g}})$ 
as $\mathrm{g}\to\infty$, we introduce a function $\Delta_{\mathrm{g}}$ of 
the coupling strength $\mathrm{g}$ such that 
there is a positive function $\delta_{\mathrm{g}}$ 
satisfying the conditions,  
\begin{eqnarray}
&{}& 
|\omega_{\mathrm{c}}-\left(\omega_{\mathrm{g}}-\Delta_{\mathrm{g}}\right)|
\le \delta_{\mathrm{g}}\omega_{\mathrm{c}}, 
\label{eq:assumption1} \\  
&{}& 
{\displaystyle \lim_{\mathrm{g}\to\infty}\delta_{\mathrm{g}}=0}.
\label{eq:assumption2}
\end{eqnarray}
These conditions yield the limit 
\begin{equation}
\lim_{\mathrm{g}\to\infty}
(\omega_{\mathrm{g}}-\Delta_{\mathrm{g}})=\omega_{\mathrm{c}}.
\label{eq:app_omega-g}
\end{equation} 
For example, take $\Delta_{\mathrm{g}}$ as $\Delta_{\mathrm{g}}=
\sqrt{\omega_{\mathrm{g}}^{2}-4\omega_{\mathrm{c}}\sqrt{\mathrm{g}C_{\mathrm{g}}
\omega_{\mathrm{c}}}}$ 
for every coupling strength $\mathrm{g}$ with 
$\sqrt{\omega_{\mathrm{c}}/\mathrm{g}C_{\mathrm{g}}}\ne 2$. 
Then, we have the equations, 
\begin{eqnarray}
&{}&
\omega_{\mathrm{c}}
\left\{
1-\frac{1}{|1-\sqrt{\omega_{\mathrm{c}}/4\mathrm{g}C_{\mathrm{g}}}|}
\right\}
=
\omega_{\mathrm{c}}
\left\{
1-\frac{1}{\sqrt{(\sqrt{\omega_{\mathrm{c}}/4\mathrm{g}C_{\mathrm{g}}})^{2}
+1^{2}-\sqrt{\omega_{\mathrm{c}}/\mathrm{g}C_{\mathrm{g}}}}}
\right\} 
\nonumber \\ 
&=&
\omega_{\mathrm{c}}
\left\{
1-\frac{1}{\sqrt{1+(\omega_{\mathrm{c}}/4\mathrm{g}C_{\mathrm{g}})
-\sqrt{\omega_{\mathrm{c}}/\mathrm{g}C_{\mathrm{g}}}}}
\right\}, 
\label{eq:eq-1}
\end{eqnarray}
and 
\begin{eqnarray}
&{}&
\omega_{\mathrm{c}}-(\omega_{\mathrm{g}}-\Delta_{\mathrm{g}}) 
\nonumber \\ 
&=& 
\omega_{\mathrm{c}}
\left\{
1-\frac{2}{
\sqrt{1+(\omega_{\mathrm{c}}/4\mathrm{g}C_{\mathrm{g}})}
+\sqrt{1+(\omega_{\mathrm{c}}/4\mathrm{g}C_{\mathrm{g}})
-\sqrt{\omega_{\mathrm{c}}/\mathrm{g}C_{\mathrm{g}}}}
}
\right\}.
\label{eq:eq-2}
\end{eqnarray}
Meanwhile, since we have the inequalities, 
\begin{eqnarray*}
&{}&
2\sqrt{1+(\omega_{\mathrm{c}}/4\mathrm{g}C_{\mathrm{g}})
-\sqrt{\omega_{\mathrm{c}}/\mathrm{g}C_{\mathrm{g}}}} \\ 
&\le& 
\sqrt{1+(\omega_{\mathrm{c}}/4\mathrm{g}C_{\mathrm{g}})}
+\sqrt{1+(\omega_{\mathrm{c}}/4\mathrm{g}C_{\mathrm{g}})
-\sqrt{\omega_{\mathrm{c}}/\mathrm{g}C_{\mathrm{g}}}} \\ 
&\le& 
2\sqrt{1+(\omega_{\mathrm{c}}/4\mathrm{g}C_{\mathrm{g}})}, 
\end{eqnarray*}
we reach the inequalities, 
\begin{eqnarray}
&{}& 
1-\frac{1}{
\sqrt{1+(\omega_{\mathrm{c}}/4\mathrm{g}C_{\mathrm{g}})
-\sqrt{\omega_{\mathrm{c}}/\mathrm{g}C_{\mathrm{g}}}}
} \nonumber \\ 
&\le& 
1-\frac{2}{
\sqrt{1+(\omega_{\mathrm{c}}/4\mathrm{g}C_{\mathrm{g}})}
+\sqrt{1+(\omega_{\mathrm{c}}/4\mathrm{g}C_{\mathrm{g}})
-\sqrt{\omega_{\mathrm{c}}/\mathrm{g}C_{\mathrm{g}}}}
} \nonumber \\ 
&\le& 
1-\frac{1}{
\sqrt{1+(\omega_{\mathrm{c}}/4\mathrm{g}C_{\mathrm{g}})}
}. 
\label{eq:ineq-1}
\end{eqnarray}
By Eqs.(\ref{eq:eq-1}), (\ref{eq:eq-2}), and (\ref{eq:ineq-1}), 
we have the following two inequalities, 
$$
\omega_{\mathrm{c}}
\left\{
1-\frac{1}{|1-\sqrt{\omega_{\mathrm{c}}/4\mathrm{g}C_{\mathrm{g}}}|}
\right\}
\le 
\omega_{\mathrm{c}}-(\omega_{\mathrm{g}}-\Delta_{\mathrm{g}}) 
\le 
\omega_{\mathrm{c}}
\left\{
1-\frac{1}{
\sqrt{1+(\omega_{\mathrm{c}}/4\mathrm{g}C_{\mathrm{g}})}
}
\right\}. 
$$
Here, we note the inequality, 
$1-\sqrt{\omega_{\mathrm{c}}/\mathrm{g}C_{\mathrm{g}}}\le 1$. 
These inequalities suggest us that 
we chose the function $\delta_{\mathrm{g}}$ as 
$$
\delta_{\mathrm{g}}=
\max
\left\{
\Biggl|
1-\frac{1}{|1-\sqrt{\omega_{\mathrm{c}}/4\mathrm{g}C_{\mathrm{g}}}|}
\Biggr|\,\,\, ,\,\,\, 
\Biggl|
1-\frac{1}{\sqrt{1+(\omega_{\mathrm{c}}/4\mathrm{g}C_{\mathrm{g}}})}
\Biggr|
\right\}. 
$$

As proved in the next section, 
we have the adiabatic approximation 
of the renormalized Hamiltonian $\mathcal{H}^{\mathrm{ren}}_{A^{2}}(\varepsilon)$: 
\begin{equation}
\mathcal{H}^{\mathrm{ren}}_{A^{2}}(\varepsilon)
\approx
U(\widetilde{\mathrm{g}}/\omega_{\mathrm{g}})^{*}
\Biggl\{
H_{\mathrm{ptn}}(\omega_{\mathrm{g}})
+\mathcal{H}_{\mathrm{atm}}(\varepsilon)
\Biggr\}
U(\widetilde{\mathrm{g}}/\omega_{\mathrm{g}})
-\hbar\frac{\widetilde{\mathrm{g}}^{2}}{\omega_{\mathrm{g}}}. 
\label{eq:main-result}
\end{equation}
We note that the whole atom energy $\mathcal{H}_{\mathrm{atm}}(\varepsilon)$ 
remains in the adiabatic approximation. 
Thanks to Theorem VIII.23(b) of Ref.\cite{RS1}, 
we can obtain the adiabatic approximation of 
the eigenstates and their corresponding eigenenergies 
in the following.

In the case $\varepsilon=0$, 
the adiabatic approximation of 
all the physical eigenstates $|\mathcal{E}_{\nu}^{\mathrm{ren}}(0)\rangle$ 
of the renormalized Hamiltonian 
$\mathcal{H}_{A^{2}}^{\mathrm{ren}}(0)$ are given by 
the same formulas of the Schr\"{o}dinger-cat-like 
entangled sates as in Eq.(\ref{eq:approximation_AN}): 
\begin{equation}
\left\{
\begin{array}{ll}
{\displaystyle \frac{1}{\sqrt{2}}
\Bigg(|\!\!\uparrow\rangle 
D\left(-\widetilde{\mathrm{g}}/\omega_{\mathrm{g}}\right)
|n\rangle
+|\!\!\downarrow\rangle 
D\left(\widetilde{\mathrm{g}}/\omega_{\mathrm{g}}\right)
|n\rangle
\Biggr)}, \\ 
\qquad \\  
{\displaystyle \frac{1}{\sqrt{2}}
\Bigg(
|\!\!\uparrow\rangle 
D\left(-\widetilde{\mathrm{g}}/\omega_{\mathrm{g}}\right)
|n\rangle
-|\!\!\downarrow\rangle 
D\left(\widetilde{\mathrm{g}}/\omega_{\mathrm{g}}\right)
|n\rangle
\Biggr)}. 
\end{array}
\right.
\label{eq:approximation_1}
\end{equation}
We denote by 
$|\mathcal{E}_{n}^{\mathrm{app},+}(0)\rangle$ 
the first expression in Eq.(\ref{eq:approximation_1}), 
and by $|\mathcal{E}_{n}^{\mathrm{app},-}(0)\rangle$ 
the second one.  
The eigenenergy of the approximated eigenstate 
$|\mathcal{E}_{n}^{\mathrm{app},+}(0)\rangle$ is 
$$
\hbar\omega_{\mathrm{g}}\left( 
n+\frac{1}{2}\right)
-\frac{\hbar\omega_{\mathrm{a}}}{2}
-\hbar\frac{\widetilde{\mathrm{g}}^{2}}{\omega_{\mathrm{g}}} 
\approx 
\hbar(\omega_{\mathrm{c}}+\Delta_{\mathrm{g}})n
+\frac{\hbar(\omega_{\mathrm{g}}-\omega_{\mathrm{a}})}{2}
-\frac{\hbar}{4C_{\infty}},
$$ 
and that of the approximated eigenstate 
$|\mathcal{E}_{n}^{\mathrm{app},-}(0)\rangle$ is 
$$
\hbar\omega_{\mathrm{g}}\left(
n+\frac{1}{2}\right)
+\frac{\hbar\omega_{\mathrm{a}}}{2}
-\hbar\frac{\widetilde{\mathrm{g}}^{2}}{\omega_{\mathrm{g}}}
\approx 
\hbar(\omega_{\mathrm{c}}+\Delta_{\mathrm{g}})n
+\frac{\hbar(\omega_{\mathrm{g}}+\omega_{\mathrm{a}})}{2}
-\frac{\hbar}{4C_{\infty}},
$$ 
which says that the tunnel splitting of the two-level atom remains in 
the adiabatic approximation. 

In the case $\varepsilon\ne 0$, 
all the physical eigenstates 
$|\mathcal{E}_{\nu}^{\mathrm{ren}}(\varepsilon)\rangle$ 
of the renormalized Hamiltonian 
$\mathcal{H}_{A^{2}}^{\mathrm{ren}}(\varepsilon)$ 
have the following adiabatic approximation:
\begin{equation}
\left\{
\begin{array}{ll}
c_{+\varepsilon,\omega_{\mathrm{a}}}
\Biggl(
-\omega_{\mathrm{a}}|\!\!\uparrow\rangle 
D(-\widetilde{\mathrm{g}}/\omega_{\mathrm{g}})|n\rangle 
+(\varepsilon-\sqrt{\varepsilon^{2}+\omega_{\mathrm{a}}^{2}})
|\!\!\downarrow\rangle 
D(\widetilde{\mathrm{g}}/\omega_{\mathrm{g}})|n\rangle
\Biggr),\\ 
\qquad \\ 
c_{-\varepsilon,\omega_{\mathrm{a}}}
\Biggl(
-\omega_{\mathrm{a}}|\!\!\uparrow\rangle 
D(-\widetilde{\mathrm{g}}/\omega_{\mathrm{g}})|n\rangle 
+(\varepsilon+\sqrt{\varepsilon^{2}+\omega_{\mathrm{a}}^{2}})
|\!\!\downarrow\rangle 
D(\widetilde{\mathrm{g}}/\omega_{\mathrm{g}})|n\rangle
\Biggr),
\end{array}
\right.
\label{eq:approximation_2}
\end{equation}
where the positive constant $c_{\pm\varepsilon,\omega_{\mathrm{a}}}$ 
is given by 
$1/c_{\pm\varepsilon,\omega_{\mathrm{a}}}^{2}=
2(\omega_{\mathrm{a}}^{2}+\varepsilon^{2}
\mp\varepsilon\sqrt{\omega_{\mathrm{a}}^{2}+\varepsilon^{2}})$. 
Eqs.(\ref{eq:approximation_2}) show up 
the Schr\"{o}dinger-cat-like entangled states. 
We denote by 
$|\mathcal{E}_{n}^{\mathrm{app},+}(\varepsilon)\rangle$ 
the first expression in Eq.(\ref{eq:approximation_2}), 
and by $|\mathcal{E}_{n}^{\mathrm{app},-}(\varepsilon)\rangle$ 
the second one.  
The eigenenergy of the adiabatically approximated eigenstates 
$|\mathcal{E}_{n}^{\mathrm{app},\pm}(\varepsilon)\rangle$ 
is 
\begin{eqnarray}
&{}& 
\mp\frac{\hbar}{2}\sqrt{\omega_{\mathrm{a}}^{2}+\varepsilon^{2}}
+\hbar\omega_{\mathrm{g}}\left(n+\frac{1}{2}\right)
-\hbar\frac{\widetilde{\mathrm{g}}^{2}}{\omega_{\mathrm{g}}}
\nonumber \\ 
&\approx& 
\mp\frac{\hbar}{2}\sqrt{\omega_{\mathrm{a}}^{2}+\varepsilon^{2}}
+\hbar(\omega_{\mathrm{c}}+\Delta_{\mathrm{g}})n
+\frac{\hbar}{2}\omega_{\mathrm{g}}
-\frac{\hbar}{4C_{\infty}}. 
\label{eq:eigenenergies_b}
\end{eqnarray}
The energy difference between the two adiabatically approximated 
eigenstates, 
$|\mathcal{E}_{n}^{\mathrm{app},-}(\varepsilon)\rangle$ and 
$|\mathcal{E}_{n}^{\mathrm{app},+}(\varepsilon)\rangle$, 
is 
$\mathcal{E}_{n}^{\mathrm{app},-}(\varepsilon)
-\mathcal{E}_{n}^{\mathrm{app},+}(\varepsilon)
=\hbar\sqrt{\omega_{\mathrm{a}}^{2}+\varepsilon^{2}}$. 
Since the energy difference between 
$\mathcal{E}_{n}^{\mathrm{app},\pm}(\varepsilon)$ and 
$\mathcal{E}_{n+1}^{\mathrm{app},\pm}(\varepsilon)$ is 
$\mathcal{E}_{n+1}^{\mathrm{app},\pm}(\varepsilon)
-\mathcal{E}_{n}^{\mathrm{app},\pm}(\varepsilon)
=\hbar\left(\omega_{\mathrm{g}}+\Delta_{\mathrm{g}}\right)$, 
we can obtain 
\begin{equation}
0<
\mathcal{E}_{n}^{\mathrm{app},-}(\varepsilon)
-\mathcal{E}_{n}^{\mathrm{app},+}(\varepsilon)
\le 
\mathcal{E}_{n+1}^{\mathrm{app},\pm}(\varepsilon)
-\mathcal{E}_{n}^{\mathrm{app},\pm}(\varepsilon)
\label{eq:energy-difference}
\end{equation}
by controlling the parameters, $\omega_{\mathrm{a}}$, 
$\varepsilon$, $\omega_{\mathrm{c}}$, $\mathrm{g}$, and $C_{\mathrm{g}}$.

We give an application example of the adiabatic approximation 
given by Eqs.(\ref{eq:approximation_1}) and 
(\ref{eq:approximation_2}). 

Take the energy-bias parameter $\varepsilon$ as 
the transverse axis now. 
The adiabatic approximation in Eq.(\ref{eq:approximation_1}) says 
that whether the energy bias is positive or negative causes 
an energy level crossing with respect to the $\varepsilon$-axis 
because the model does not have the $A^{2}$-term. 
On the other hand, the adiabatic-approximation formula, 
Eq.(\ref{eq:approximation_2}) with Eq.(\ref{eq:eigenenergies_b}), 
says that the $A^{2}$-term makes an avoided crossing 
with respect to the $\varepsilon$-axis.   

In the same way as the proof of Eq.(21) of Ref.\cite{HMS}, 
we can obtain the expression of 
the ground-state expectation $N^{\mathrm{ren}}_{0}
:=\langle\mathcal{E}^{\mathrm{ren}}_{0}|a^{\dagger}a
|\mathcal{E}^{\mathrm{ren}}_{0}\rangle$ 
of the number of dressed photons and estimate it as   
\begin{eqnarray*}
N_{0}^{\mathrm{ren}}
&=&
\hbar^{2}\widetilde{\mathrm{g}}^{2}
\sum_{\nu=0}^{\infty}
\frac{|\langle\mathcal{E}_{\nu}^{\mathrm{ren}}|
\sigma_{z}|\mathcal{E}_{0}^{\mathrm{ren}}\rangle|^{2}}{
(E_{\nu}-E_{0}+\hbar\omega_{\mathrm{g}})^{2}
} \\ 
&\le& 
\frac{\widetilde{\mathrm{g}}^{2}}{\omega_{\mathrm{g}}^{2}}
\sum_{\nu=0}^{\infty}
|\langle\mathcal{E}_{\nu}^{\mathrm{ren}}|
\sigma_{z}|\mathcal{E}_{0}^{\mathrm{ren}}\rangle|^{2}
=
\frac{\widetilde{\mathrm{g}}^{2}}{\omega_{\mathrm{g}}^{2}}
\|\sigma_{z}|\mathcal{E}_{0}^{\mathrm{ren}}\rangle
\|_{\mathbb{C}^{2}\otimes L^{2}(\mathbb{R})}^{2}
\le
\frac{\widetilde{\mathrm{g}}^{2}}{\omega_{\mathrm{a}}^{2}}.
\end{eqnarray*}
We define the approximated ground-state expectation 
$N_{0}^{\mathrm{app}}$ 
using the adiabatic approximation, 
i.e.,  
$N_{0}^{\mathrm{app}}:=
\langle\mathcal{E}_{0}^{\mathrm{app},+}(\varepsilon)|
a^{\dagger}a|\mathcal{E}_{0}^{\mathrm{app},+}(\varepsilon)\rangle$. 
Then, the immediate calculation gives us 
the expression, 
\begin{equation}
N_{0}^{\mathrm{app}}=
\frac{\widetilde{\mathrm{g}}^{2}}{\omega_{\mathrm{g}}^{2}}
=\omega_{\mathrm{c}}^{-1/2}
\left(
\omega_{\mathrm{c}}\mathrm{g}^{-4/3}
+4\mathrm{g}^{-1/3}C_{\mathrm{g}}
\right)^{-3/2},
\label{eq:adiabatic-approximation_N_0}
\end{equation}
and thus, we have the limit, 
$\lim_{\mathrm{g}\to\infty}N_{0}^{\mathrm{ren}}=
\lim_{\mathrm{g}\to\infty}N_{0}^{\mathrm{app}}=0$, 
with $N_{0}^{\mathrm{ren}}\le N_{0}^{\mathrm{app}}$. 
Meanwhile, for the one-mode photon field $\Phi^{\mathrm{ren}}
:=(a+a^{\dagger})/\sqrt{2\omega_{\mathrm{g}}}$ 
our adiabatic approximation immediately shows 
that the fluctuation $\Delta\Phi^{\mathrm{ren}}$ decays 
to zero as $\mathrm{g}\to\infty$. 

We have information on the dressed photon 
in the ground state in the following. 
The symbol $\sharp$ stands for one of the (in)equality symbols, 
$>$, $=$, $<$. 
Eq.(\ref{eq:adiabatic-approximation_N_0}) says that 
the (in)equality, $N_{0}^{\mathrm{app}}\,\sharp\, 1$, 
is equivalent to the 
(in)equality,
\begin{equation}
\frac{1}{4}\left(\frac{\mathrm{g}}{\omega_{\mathrm{c}}}\right)^{-1}
\left\{
\left(\frac{\mathrm{g}}{\omega_{\mathrm{c}}}\right)^{4/3}-1
\right\}\,\,\,
\sharp\,\,\, C_{\mathrm{g}}.
\label{eq:equivalence1}
\end{equation}
Particularly, in the case where the function $C_{\mathrm{g}}$ 
is given by $C_{\mathrm{g}}=C\mathrm{g}$ with a constant $C$,  
Eq.(\ref{eq:equivalence1}) can be written as 
\begin{equation}
\frac{1}{4\omega_{\mathrm{c}}}
\left(\frac{\mathrm{g}}{\omega_{\mathrm{c}}}\right)^{-2}
\left\{
\left(\frac{\mathrm{g}}{\omega_{\mathrm{c}}}\right)^{4/3}-1
\right\}\,\,\,
\sharp\,\,\, C.
\label{eq:equivalence2}
\end{equation}
Namely, following the (meson) pair theory, 
if we can make the constant $C$ so small that 
it satisfies the condition, $(\omega_{\mathrm{c}}/4\mathrm{g}^{2})
\{(\mathrm{g}/\omega_{\mathrm{c}})^{4/3}-1\}\ge C$, 
then there is a possibility that the ground state of the 
generalized quantum Rabi model has some dressed photons. 
We will explain the reason why we are interested in 
Eqs.(\ref{eq:equivalence1}) and (\ref{eq:equivalence2}) 
in Section \ref{sec:CD}. 

\section{A Proof of Adiabatic Approximation for 
$\mathcal{H}^{\mathrm{ren}}_{A^{2}}(\varepsilon)$}
\label{sec:proof}

We define the modified photon Hamiltonian 
$\widetilde{H}_{\mathrm{ptn}}(\omega)$ 
for every frequency $\omega$ 
by removing the zero-point energy, that is, 
$$
\widetilde{H}_{\mathrm{ptn}}(\omega)
:=H_{\mathrm{ptn}}(\omega)
-\hbar\omega/2
=\hbar\omega a^{\dagger}a.
$$ 
In this section, we will take 
frequencies, $\omega_{\mathrm{c}}$, $\omega_{\mathrm{g}}$, 
$\omega_{\mathrm{c}}-(\omega_{\mathrm{g}}-\Delta_{\mathrm{g}})$, 
or $\omega_{\mathrm{c}}-\omega_{\mathrm{g}}$ 
as $\omega$. 
With this modification, we slightly modify the renormalized 
Hamiltonian $\mathcal{H}_{A^{2}}^{\mathrm{ren}}(\varepsilon)$ as 
$\widetilde{\mathcal{H}}_{A^{2}}^{\mathrm{ren}}(\varepsilon)
:=\mathcal{H}_{A^{2}}^{\mathrm{ren}}-\hbar\omega_{\mathrm{g}}/2$.
By Eq.(\ref{eq:renormalized-Hamiltonian}), we have the expression,  
$$
\widetilde{\mathcal{H}}_{A^{2}}^{\mathrm{ren}}(\varepsilon)
=\mathcal{H}_{\mbox{\tiny GQR}}(\omega_{\mathrm{g}},\widetilde{\mathrm{g}})
-\hbar\omega_{\mathrm{g}}/2
=\mathcal{H}_{\mathrm{atm}}(\varepsilon)
+\widetilde{H}_{\mathrm{ptn}}(\omega_{\mathrm{g}})+
\hbar\widetilde{\mathrm{g}}\sigma_{z}(a+a^{\dagger}).
$$ 
All the eigenstates of the slightly modified 
Hamiltonian 
$\widetilde{\mathcal{H}}_{A^{2}}^{\mathrm{ren}}(\varepsilon)$ 
are completely same as 
those of the original Hamiltonian 
$\mathcal{H}_{A^{2}}^{\mathrm{ren}}(\varepsilon)$. 
Thus, we prove our desired results for the Hamiltonian 
$\widetilde{\mathcal{H}}_{A^{2}}^{\mathrm{ren}}(\varepsilon)$. 

Correspondingly, we denote the modified 
free Hamiltonian $\widetilde{\mathcal{H}}_{0}$ 
of the atom-photon system 
by 
$$\widetilde{\mathcal{H}}_{0}:=\mathcal{H}_{\mathrm{atm}}(\varepsilon)
+\widetilde{H}_{\mathrm{ptn}}(\omega_{\mathrm{c}}).
$$ 
All the eigenenergies of the atom Hamiltonian 
$\mathcal{H}_{\mathrm{atm}}(\varepsilon)$ 
are $\pm(\hbar/2)\sqrt{\omega_{\mathrm{a}}^{2}+\varepsilon^{2}}$, 
and thus, we have its operator norm, 
$\|\mathcal{H}_{\mathrm{atm}}(\varepsilon)\|_{\mathrm{op}}
=(\hbar/2)\sqrt{\omega_{\mathrm{a}}^{2}+\varepsilon^{2}}$. 
Since we can rewrite the Hamiltonian 
$\widetilde{H}_{\mathrm{ptn}}(\omega_{\mathrm{c}})$ 
as $\widetilde{H}_{\mathrm{ptn}}(\omega_{\mathrm{c}})
=\widetilde{\mathcal{H}}_{0}-\mathcal{H}_{\mathrm{atm}}(\varepsilon)$, 
we have 
\begin{eqnarray*}
\|\widetilde{H}_{\mathrm{ptn}}(\omega_{\mathrm{c}})\Psi
\|_{\mathbb{C}^{2}\otimes L^{2}(\mathbb{R})}
&\le& 
\|\widetilde{\mathcal{H}}_{0}\Psi
\|_{\mathbb{C}^{2}\otimes L^{2}(\mathbb{R})}
+\|\mathcal{H}_{\mathrm{atm}}(\varepsilon)\|_{\mathrm{op}}
\|\Psi\|_{\mathbb{C}^{2}\otimes L^{2}(\mathbb{R})} \\ 
&=&
\|\widetilde{\mathcal{H}}_{0}\Psi
\|_{\mathbb{C}^{2}\otimes L^{2}(\mathbb{R})}
+\frac{\hbar}{2}\sqrt{\omega_{\mathrm{a}}^{2}+\varepsilon^{2}}
\|\Psi\|_{\mathbb{C}^{2}\otimes L^{2}(\mathbb{R})} 
\end{eqnarray*}
for every vector $\Psi\in\mathbb{C}^{2}\otimes L^{2}(\mathbb{R})$.
In particular, we set 
$\Psi=(\widetilde{\mathcal{H}}_{0}-i\hbar)^{-1}\Phi$ 
for every vector $\Phi\in\mathbb{C}^{2}\otimes L^{2}(\mathbb{R})$, 
and insert it into the above inequality. 
Then, we have the inequalities, 
\begin{eqnarray}
&{}& 
\|\widetilde{H}_{\mathrm{ptn}}(\omega_{\mathrm{c}})
(\widetilde{\mathcal{H}}_{0}-i\hbar)^{-1}
\Phi
\|_{\mathbb{C}^{2}\otimes L^{2}(\mathbb{R})} 
\nonumber \\ 
&\le& 
\|\widetilde{\mathcal{H}}_{0}
(\widetilde{\mathcal{H}}_{0}-i\hbar)^{-1}
\Phi
\|_{\mathbb{C}^{2}\otimes L^{2}(\mathbb{R})} 
+\frac{\hbar}{2}\sqrt{\omega_{\mathrm{a}}^{2}+\varepsilon^{2}}
\|
(\widetilde{\mathcal{H}}_{0}-i\hbar)^{-1}
\Phi
\|_{\mathbb{C}^{2}\otimes L^{2}(\mathbb{R})} 
\nonumber \\ 
&\le& 
\|\widetilde{\mathcal{H}}_{0}
(\widetilde{\mathcal{H}}_{0}-i\hbar)^{-1}\|_{\mathrm{op}}
\|\Phi
\|_{\mathbb{C}^{2}\otimes L^{2}(\mathbb{R})} 
+\frac{\hbar}{2}\sqrt{\omega_{\mathrm{a}}^{2}+\varepsilon^{2}}
\|(\widetilde{\mathcal{H}}_{0}-i\hbar)^{-1}\|_{\mathrm{op}}
\|\Phi
\|_{\mathbb{C}^{2}\otimes L^{2}(\mathbb{R})}.
\label{eq:proof-1_a}
\end{eqnarray}
To estimate the operator norms, 
$\|\widetilde{\mathcal{H}}_{0}
(\widetilde{\mathcal{H}}_{0}-i\hbar)^{-1}\|_{\mathrm{op}}$ 
and 
$\|(\widetilde{\mathcal{H}}_{0}-i\hbar)^{-1}\|_{\mathrm{op}}$, 
we make a general argument. 
Let $\mathcal{H}$ be an arbitrary self-adjoint energy operator 
(i.e., Hamiltonian). 
We recall Theorem VIII.6 of Ref.\cite{RS1} 
or Theorems 7.14 and 7.17 of Ref.\cite{weidmann}:  
There is a spectral family 
(i.e., the set of projection-valued measures) $P_{\xi}^{\mathcal{H}}$
for the Hamiltonian $\mathcal{H}$ 
so that  
$$
\mathcal{H}=
\int_{-\infty}^{\infty}\xi\, 
dP_{\xi}^{\mathcal{H}}. 
$$
Using the properties of the projection-valued measures 
$P_{\xi}^{\mathcal{H}}$, 
we have the following estimates, 
\begin{eqnarray*}
\|\mathcal{H}
(\mathcal{H}-i\hbar)^{-1}
\Phi
\|_{\mathbb{C}^{2}\otimes L^{2}(\mathbb{R})}^{2}
&=&
\int_{-\infty}^{\infty}
\left|
\frac{\xi}{\xi-i\hbar}
\right|^{2}
d\|
P_{\xi}^{\mathcal{H}}\Phi
\|_{\mathbb{C}^{2}\otimes L^{2}(\mathbb{R})}^{2} \\ 
&\le& 
\int_{-\infty}^{\infty}
d\|
P_{\xi}^{\mathcal{H}}\Phi
\|_{\mathbb{C}^{2}\otimes L^{2}(\mathbb{R})}^{2}
=\|\Phi\|_{\mathbb{C}^{2}\otimes L^{2}(\mathbb{R})}^{2}
\end{eqnarray*}
and 
\begin{eqnarray*}
\|(\mathcal{H}-i\hbar)^{-1}
\Phi
\|_{\mathbb{C}^{2}\otimes L^{2}(\mathbb{R})}^{2}
&=&
\int_{-\infty}^{\infty}
\left|
\frac{1}{\xi-i\hbar}
\right|^{2}
d\|
P_{\xi}^{\mathcal{H}}\Phi
\|_{\mathbb{C}^{2}\otimes L^{2}(\mathbb{R})}^{2} \\ 
&\le& 
\frac{1}{\hbar^{2}}
\int_{-\infty}^{\infty}
d\|
P_{\xi}^{\mathcal{H}}\Phi
\|_{\mathbb{C}^{2}\otimes L^{2}(\mathbb{R})}^{2}
=\frac{1}{\hbar^{2}}\|\Phi\|_{\mathbb{C}^{2}\otimes L^{2}(\mathbb{R})}^{2},  
\end{eqnarray*}
for every vector $\Phi\in\mathbb{C}^{2}\otimes L^{2}(\mathbb{R})$. 
These estimates bring us the two operator-norm inequalities, 
$$
\|\mathcal{H}(\mathcal{H}-i\hbar)^{-1}\|_{\mathrm{op}}
\le 1\quad 
\mbox{and}\quad 
\|(\mathcal{H}-i\hbar)^{-1}\|_{\mathrm{op}}
\le1/\hbar. 
$$
Inserting the inequalities,
$\|\widetilde{\mathcal{H}}_{0}
(\widetilde{\mathcal{H}}_{0}-i\hbar)^{-1}\|_{\mathrm{op}}
\le 1$ and $\|(\widetilde{\mathcal{H}}_{0}-i\hbar)^{-1}\|_{\mathrm{op}}
\le1/\hbar$, 
into Eq.(\ref{eq:proof-1_a}), 
we reach the inequality 
\begin{equation}
\|\widetilde{H}_{\mathrm{ptn}}(\omega_{\mathrm{c}})
(\widetilde{\mathcal{H}}_{0}-i\hbar)^{-1}
\|_{\mathrm{op}}
\le 
1+\frac{1}{2}\sqrt{\omega_{\mathrm{a}}^{2}+\varepsilon^{2}}.
\label{eq:proof-1_b}
\end{equation}
Meanwhile, since we assume Eq.(\ref{eq:assumption1}), we have 
\begin{eqnarray*}
&{}& 
\|\widetilde{H}_{\mathrm{ptn}}\left(
\omega_{\mathrm{c}}
-\left(\omega_{\mathrm{g}}-\Delta_{\mathrm{g}}\right)
\right)
(\widetilde{\mathcal{H}}_{0}-i\hbar)^{-1}
\Phi
\|_{\mathbb{C}^{2}\otimes L^{2}(\mathbb{R})} \\ 
&=&
\hbar |\omega_{\mathrm{c}}
-\left(\omega_{\mathrm{g}}-\Delta_{\mathrm{g}}\right)|\,
\| a^{\dagger}a
(\widetilde{\mathcal{H}}_{0}-i\hbar)^{-1}
\Phi
\|_{\mathbb{C}^{2}\otimes L^{2}(\mathbb{R})} \\ 
&\le& 
\hbar\delta_{\mathrm{g}}\omega_{\mathrm{c}}
\| a^{\dagger}a
(\widetilde{\mathcal{H}}_{0}-i\hbar)^{-1}
\Phi
\|_{\mathbb{C}^{2}\otimes L^{2}(\mathbb{R})} \\ 
&=& 
\delta_{\mathrm{g}}
\|\widetilde{H}_{\mathrm{ptn}}(\omega_{\mathrm{c}})
(\widetilde{\mathcal{H}}_{0}-i\hbar)^{-1}
\Phi
\|_{\mathbb{C}^{2}\otimes L^{2}(\mathbb{R})},
\end{eqnarray*}
which implies the operator-norm inequality, 
\begin{equation}
\|\widetilde{H}_{\mathrm{ptn}}\left(
\omega_{\mathrm{c}}
-\left(\omega_{\mathrm{g}}-\Delta_{\mathrm{g}}\right)
\right)
(\widetilde{\mathcal{H}}_{0}-i\hbar)^{-1}
\|_{\mathrm{op}}
\le
\delta_{\mathrm{g}}
\|\widetilde{H}_{\mathrm{ptn}}(\omega_{\mathrm{c}})
(\widetilde{\mathcal{H}}_{0}-i\hbar)^{-1}
\|_{\mathrm{op}}. 
\label{eq:proof-1_c}
\end{equation}
Combining Eqs.(\ref{eq:proof-1_b}) and (\ref{eq:proof-1_c}), 
we reach the inequality, 
\begin{equation}
\|\widetilde{H}_{\mathrm{ptn}}(
\omega_{\mathrm{c}}
-(\omega_{\mathrm{g}}-\Delta_{\mathrm{g}}))
(\widetilde{\mathcal{H}}_{0}-i\hbar)^{-1}
\|_{\mathrm{op}}
\le 
\delta_{\mathrm{g}}
\left(
1+\frac{1}{2}\sqrt{\omega_{\mathrm{a}}^{2}+\varepsilon^{2}}
\right).
\label{eq:proof-1}
\end{equation}

Let $G$ be $\widetilde{\mathrm{g}}/\omega_{\mathrm{g}}$, i.e., 
$G:=\widetilde{\mathrm{g}}/\omega_{\mathrm{g}}$. 
Then, we immediately know that 
this quantity $G$ decays to the zero as the coupling strength $\mathrm{g}$ 
tends to the infinity, i.e., 
$\lim_{\mathrm{g}\to\infty}G=0$. 
For simplicity, we denote 
the displacement operators 
$D(\pm\widetilde{\mathrm{g}}/\omega_{\mathrm{g}})$
by $D_{\pm}(G)$, i.e., 
$D_{\pm}(G):=
D(\pm\widetilde{\mathrm{g}}/\omega_{\mathrm{g}})$. 
We know the expression, $D_{\pm}(G)=e^{\pm iG\{i(a-a^{\dagger})\}}$. 
Since the operator $i(a-a^{\dagger})$ 
is self-adjoint on the domain 
of the photon number operator $a^{\dagger}a$, 
the operator $D_{\pm}(G)=e^{\pm iG\{i(a-a^{\dagger})\}}$ 
is a strongly continuous one-parameter unitary group 
by Theorem VIII.7 of Ref.\cite{RS1}. 
Namely, we have the limit, 
$\lim_{G\to 0}(1-D_{\pm}(G)^{2})\Psi=0$, 
for every vector $\Psi\in\mathbb{C}^{2}\otimes L^{2}(\mathbb{R})$. 
Thus, the operator $\sigma_{+}(1-D_{+}(G)^{2})+
\sigma_{-}(1-D_{-}(G)^{2})$ goes to the zero operator 
as $\mathrm{g}\to\infty$ in the strong operator topology, 
and therefore, 
in the weak operator topology. 
For details on these topologies, see \S VI.1 of Ref.\cite{RS1}. 
We note that the inequality, 
$\|\sigma_{+}(1-D_{+}(G)^{2})+
\sigma_{-}(1-D_{-}(G)^{2})\|_{\mathrm{op}}\le 2$, holds, 
and that the resolvent 
$(\widetilde{\mathcal{H}}_{0}-i\hbar)^{-1}$ is compact. 
Therefore, by applying Theorem in Appendix A of Ref.\cite{HMS}, 
we obtain the limit, 
\begin{equation}
\lim_{\mathrm{g}\to\infty}
\Biggl\|
(\widetilde{\mathcal{H}}_{0}-i\hbar)^{-1}
\Bigl\{\sigma_{+}\left(1-D_{+}(G)^{2}\right)
+\sigma_{-}\left(1-D_{-}(G)^{2}\right)\Bigr\}
(\widetilde{\mathcal{H}}_{0}-i\hbar)^{-1}
\Biggr\|_{\mathrm{op}}=0.
\label{eq:proof-2}
\end{equation}
From now on, we denote the operator 
$\sigma_{+}D_{+}(G)^{2}
+\sigma_{-}D_{-}(G)^{2}$ 
by $\Xi_{0}(\mathrm{g})$, 
and moreover, 
the operator $\sigma_{x}-\Xi_{0}(\mathrm{g})$ 
by $\Xi_{1}(\mathrm{g})$: 
\begin{eqnarray*}
\Xi_{0}(\mathrm{g})&:=&
\sigma_{+}D_{+}(G)^{2}+\sigma_{-}D_{-}(G)^{2}, \\ 
\Xi_{1}(\mathrm{g})&:=&
\sigma_{x}-\Xi_{0}(\mathrm{g})
=\sigma_{+}\left(1-D_{+}(G)^{2}\right)
+\sigma_{-}\left(1-D_{-}(G)^{2}\right). 
\end{eqnarray*}
Moreover, we define a Hamiltonian 
$\widetilde{\mathcal{H}}(\mathrm{g})$ by 
$$
\widetilde{\mathcal{H}}(\mathrm{g}):=
\widetilde{H}_{\mathrm{ptn}}(\omega_{\mathrm{g}}-\Delta_{\mathrm{g}})
-\frac{\hbar}{2}\varepsilon\sigma_{z}
-\frac{\hbar}{2}\omega_{\mathrm{a}}\Xi_{0}(\mathrm{g}).
$$
We define an operator $R$ by 
the difference between the resolvents 
of $\widetilde{\mathcal{H}}(\mathrm{g})$ 
and $\widetilde{\mathcal{H}}_{0}$, 
that is, 
$$
R:=
(\widetilde{\mathcal{H}}(\mathrm{g})-i\hbar)^{-1}
-
(\widetilde{\mathcal{H}}_{0}-i\hbar)^{-1}. 
$$
Using the $2$nd resolvent identity in 
Theorem 5.13(b) of Ref.\cite{weidmann} 
and the equation $\widetilde{H}_{\mathrm{ptn}}(\omega_{\mathrm{c}})
-\widetilde{H}_{\mathrm{ptn}}(\omega_{\mathrm{g}}-\Delta_{\mathrm{g}})
=\widetilde{H}_{\mathrm{ptn}}(\omega_{\mathrm{c}}-
(\omega_{\mathrm{g}}-\Delta_{\mathrm{g}}))$, 
we can calculate the expression of 
the difference operator $R$ as 
\begin{eqnarray}
R&=& 
(\widetilde{\mathcal{H}}(\mathrm{g})-i\hbar)^{-1}
\Bigl\{
\widetilde{H}_{\mathrm{ptn}}(\omega_{\mathrm{c}}-(\omega_{\mathrm{g}}-\Delta_{\mathrm{g}}))
-\frac{\hbar}{2}\omega_{\mathrm{a}}\Xi_{1}(\mathrm{g}) 
\Bigr\}
(\widetilde{\mathcal{H}}_{0}-i\hbar)^{-1}. 
\label{eq:proof-a}
\end{eqnarray}
Insert this into the equation 
$(\widetilde{\mathcal{H}}(\mathrm{g})-i\hbar)^{-1}
=(\widetilde{\mathcal{H}}_{0}-i\hbar)^{-1}+R$, 
then we have the equation, 
\begin{eqnarray}
(\widetilde{\mathcal{H}}(\mathrm{g})-i\hbar)^{-1}
&=& 
(\widetilde{\mathcal{H}}_{0}-i\hbar)^{-1}  
+
(\widetilde{\mathcal{H}}(\mathrm{g})-i\hbar)^{-1}
\widetilde{H}_{\mathrm{ptn}}
(\omega_{\mathrm{c}}-(\omega_{\mathrm{g}}-\Delta_{\mathrm{g}}))
(\widetilde{\mathcal{H}}_{0}
-i\hbar)^{-1} 
\nonumber \\ 
&{}&
-\frac{\hbar}{2}\omega_{\mathrm{a}}
(\widetilde{\mathcal{H}}(\mathrm{g})-i\hbar)^{-1}
\Xi_{1}(\mathrm{g})
(\widetilde{\mathcal{H}}_{0}-i\hbar)^{-1}. 
\label{eq:proof-b}
\end{eqnarray}
Inserting Eq.(\ref{eq:proof-b}) into Eq.(\ref{eq:proof-a}), 
we have the decomposition, 
\begin{eqnarray*}
R&=& 
(\widetilde{\mathcal{H}}_{0}-i\hbar)^{-1}
\left\{
\widetilde{H}_{\mathrm{ptn}}(\omega_{\mathrm{c}}-
(\omega_{\mathrm{g}}-\Delta_{\mathrm{g}}))
-\frac{\hbar\omega_{\mathrm{a}}}{2}\Xi_{1}(\mathrm{g})
\right\}(\widetilde{\mathcal{H}}_{0}-i\hbar)^{-1} \\  
&{}&
+(\widetilde{\mathcal{H}}(\mathrm{g})-i\hbar)^{-1}
\widetilde{H}_{\mathrm{ptn}}(\omega_{\mathrm{c}}-
(\omega_{\mathrm{g}}-\Delta_{\mathrm{g}}))
(\widetilde{\mathcal{H}}_{0}-i\hbar)^{-1} \\ 
&{}&\qquad\qquad\quad 
\left\{
\widetilde{H}_{\mathrm{ptn}}(\omega_{\mathrm{c}}-
(\omega_{\mathrm{g}}-\Delta_{\mathrm{g}}))
-\frac{\hbar\omega_{\mathrm{a}}}{2}\Xi_{1}(\mathrm{g})
\right\}(\widetilde{\mathcal{H}}_{0}-i\hbar)^{-1} \\ 
&{}&
-\frac{\hbar\omega_{\mathrm{a}}}{2}
(\widetilde{\mathcal{H}}(\mathrm{g})-i\hbar)^{-1}
\Xi_{1}(\mathrm{g})
(\widetilde{\mathcal{H}}_{0}-i\hbar)^{-1} \\ 
&{}&\qquad\qquad\quad 
\left\{
\widetilde{H}_{\mathrm{ptn}}(\omega_{\mathrm{c}}-
(\omega_{\mathrm{g}}-\Delta_{\mathrm{g}}))
-\frac{\hbar\omega_{\mathrm{a}}}{2}\Xi_{1}(\mathrm{g})
\right\}(\widetilde{\mathcal{H}}_{0}-i\hbar)^{-1}. 
\end{eqnarray*}
Eventually, we can decompose the difference operator $R$ as 
\begin{equation}
R=\sum_{j=1}^{6}I_{j},\,\,\, 
\mbox{and thus},\,\,\, 
\|R\|_{\mathrm{op}}=\sum_{j=1}^{6}\|I_{j}\|_{\mathrm{op}},
\label{eq:proof-c}
\end{equation}
where 
\begin{eqnarray*}
I_{1}&=&
(\widetilde{\mathcal{H}}_{0}-i\hbar)^{-1}
\widetilde{H}_{\mathrm{ptn}}(
\omega_{\mathrm{c}}-(\omega_{\mathrm{g}}-\Delta_{\mathrm{g}}))
(\widetilde{\mathcal{H}}_{0}-i\hbar)^{-1}, \\ 
I_{2}&=&
-\frac{\hbar\omega_{\mathrm{a}}}{2}
(\widetilde{\mathcal{H}}_{0}-i\hbar)^{-1}
\Xi_{1}(\mathrm{g}) 
(\widetilde{\mathcal{H}}_{0}-i\hbar)^{-1}, \\ 
I_{3}&=&
(\widetilde{\mathcal{H}}(\mathrm{g})-i\hbar)^{-1}
\Bigl\{
\widetilde{H}_{\mathrm{ptn}}(\omega_{\mathrm{c}}-(\omega_{\mathrm{g}}-\Delta_{\mathrm{g}})) 
(\widetilde{\mathcal{H}}_{0}-i\hbar)^{-1}
\Bigr\}^{2}, \\ 
I_{4}&=&
-\frac{\hbar\omega_{\mathrm{a}}}{2}
(\widetilde{\mathcal{H}}(\mathrm{g})-i\hbar)^{-1}
\widetilde{H}_{\mathrm{ptn}}(\omega_{\mathrm{c}}-(\omega_{\mathrm{g}}-\Delta_{\mathrm{g}}))
(\widetilde{\mathcal{H}}_{0}-i\hbar)^{-1} 
\Xi_{1}(\mathrm{g})
(\widetilde{\mathcal{H}}_{0}-i\hbar)^{-1}, \\ 
I_{5}&=&
-\frac{\hbar\omega_{\mathrm{a}}}{2}
(\widetilde{\mathcal{H}}(\mathrm{g})-i\hbar)^{-1}
\Xi_{1}(\mathrm{g})
(\widetilde{\mathcal{H}}_{0}-i\hbar)^{-1}
\widetilde{H}_{\mathrm{ptn}}(\omega_{\mathrm{c}}-(\omega_{\mathrm{g}}-\Delta_{\mathrm{g}}))
(\widetilde{\mathcal{H}}_{0}-i\hbar)^{-1},  \\ 
I_{6}&=&
\left(-\, \frac{\hbar\omega_{\mathrm{a}}}{2}\right)^{2}
(\widetilde{\mathcal{H}}(\mathrm{g})-i\hbar)^{-1}
\Xi_{1}(\mathrm{g})
(\widetilde{\mathcal{H}}_{0}-i\hbar)^{-1}
\Xi_{1}(\mathrm{g})
(\widetilde{\mathcal{H}}_{0}-i\hbar)^{-1}.  
\end{eqnarray*}
Using Eq.(\ref{eq:proof-1}) and the inequalities, 
$\|\widetilde{\mathcal{H}}_{0}-i\hbar\|_{\mathrm{op}}
\le 1/\hbar$,  
$\|\widetilde{\mathcal{H}}(\mathrm{g})-i\hbar\|_{\mathrm{op}}
\le 1/\hbar$, $\|\Xi_{1}(\mathrm{g})\|_{\mathrm{op}}\le 2$, 
individual operators $I_{j}$ are bounded from above 
in the following: 
\begin{eqnarray*}
\|I_{1}\|_{\mathrm{op}}&\le&
\|(\widetilde{\mathcal{H}}_{0}-i\hbar)^{-1}\|_{\mathrm{op}}
\|\widetilde{H}_{\mathrm{ptn}}(
\omega_{\mathrm{c}}-(\omega_{\mathrm{g}}-\Delta_{\mathrm{g}}))
(\widetilde{\mathcal{H}}_{0}-i\hbar)^{-1}
\|_{\mathrm{op}} \\ 
&\le&
\frac{\delta_{\mathrm{g}}}{\hbar}
\left(
1+\frac{1}{2}\sqrt{\omega_{\mathrm{a}}^{2}+\varepsilon^{2}}
\right), 
\end{eqnarray*}
\begin{eqnarray*}
\|I_{2}\|_{\mathrm{op}} 
&\le&
\frac{\hbar\omega_{\mathrm{a}}}{2}
\|(\widetilde{\mathcal{H}}_{0}-i\hbar)^{-1}
\Xi_{1}(\mathrm{g})
(\widetilde{\mathcal{H}}_{0}-i\hbar)^{-1}
\|_{\mathrm{op}}, 
\end{eqnarray*}
\begin{eqnarray*}
\|I_{3}\|_{\mathrm{op}}
&\le&
\|(\widetilde{\mathcal{H}}(\mathrm{g})-i\hbar)^{-1}
\|_{\mathrm{op}}
\|
\widetilde{H}_{\mathrm{ptn}}(\omega_{\mathrm{c}}-(\omega_{\mathrm{g}}-\Delta_{\mathrm{g}})) 
(\widetilde{\mathcal{H}}_{0}-i\hbar)^{-1}
\|_{\mathrm{op}}^{2} \\ 
&\le&
\frac{\delta_{\mathrm{g}}^{2}}{\hbar}
\left(
1+\frac{1}{2}\sqrt{\omega_{\mathrm{a}}^{2}+\varepsilon^{2}}
\right)^{2}, 
\end{eqnarray*}
\begin{eqnarray*}
\|I_{4}\|_{\mathrm{op}}
&\le& 
\frac{\hbar}{2}\omega_{\mathrm{a}}
\|(\widetilde{\mathcal{H}}(\mathrm{g})-i\hbar)^{-1}\|_{\mathrm{op}}
\|\widetilde{H}_{\mathrm{ptn}}(\omega_{\mathrm{c}}
-(\omega_{\mathrm{g}}-\Delta_{\mathrm{g}}))
(\widetilde{\mathcal{H}}_{0}-i\hbar)^{-1}\|_{\mathrm{op}} \\ 
&{}&\qquad\qquad\times
\|\Xi_{1}(\mathrm{g})\|_{\mathrm{op}}
\|(\widetilde{\mathcal{H}}_{0}-i\hbar)^{-1}\|_{\mathrm{op}} \\ 
&\le&
\frac{\omega_{\mathrm{a}}\delta_{\mathrm{g}}}{\hbar}
\left(
1+\frac{1}{2}\sqrt{\omega_{\mathrm{a}}^{2}+\varepsilon^{2}}
\right), \\ 
\|I_{5}\|_{\mathrm{op}}
&\le&
\frac{\hbar}{2}\omega_{\mathrm{a}}
\|(\widetilde{\mathcal{H}}(\mathrm{g})-i\hbar)^{-1}\|_{\mathrm{op}}
\|\Xi_{1}(\mathrm{g})\|_{\mathrm{op}}
\|(\widetilde{\mathcal{H}}_{0}-i\hbar)^{-1}\|_{\mathrm{op}} \\ 
&{}&\qquad\qquad\times 
\|\widetilde{H}_{\mathrm{ptn}}(\omega_{\mathrm{c}}
-(\omega_{\mathrm{g}}-\Delta_{\mathrm{g}}))
(\widetilde{\mathcal{H}}_{0}-i\hbar)^{-1}\|_{\mathrm{op}} \\ 
&\le&
\frac{\omega_{\mathrm{a}}\delta_{\mathrm{g}}}{\hbar}
\left(
1+\frac{1}{2}\sqrt{\omega_{\mathrm{a}}^{2}+\varepsilon^{2}}
\right), \\ 
\|I_{6}\|_{\mathrm{op}}
&\le& 
\frac{\hbar^{2}\omega_{\mathrm{a}}^{2}}{4}
\|(\widetilde{\mathcal{H}}(\mathrm{g})-i\hbar)^{-1}\|_{\mathrm{op}}
\|\Xi_{1}(\mathrm{g})\|_{\mathrm{op}}
\|(\widetilde{\mathcal{H}}_{0}-i\hbar)^{-1}
\Xi_{1}(\mathrm{g})
(\widetilde{\mathcal{H}}_{0}-i\hbar)^{-1}\|_{\mathrm{op}} \\ 
&\le&
\frac{\hbar\omega_{\mathrm{a}}^{2}}{2}
\|(\widetilde{\mathcal{H}}_{0}-i\hbar)^{-1}
\Xi_{1}(\mathrm{g})
(\widetilde{\mathcal{H}}_{0}-i\hbar)^{-1}
\|_{\mathrm{op}}. 
\end{eqnarray*}
Eqs.(\ref{eq:assumption2}) and (\ref{eq:proof-2}) 
tell us that all the operator norms of 
operators $I_{j}$ converges to zero as taking strong coupling 
limit, i.e., 
$\lim_{\mathrm{g}\to\infty}\| I_{j}\|_{\mathrm{op}}=0$. 
Therefore, by Eq.(\ref{eq:proof-c}), 
we obtain our desired limit, 
$\lim_{\mathrm{g}\to\infty}\| R\|_{\mathrm{op}}=0$. 
Namely, we succeed in proving the convergence,
\begin{equation}
\widetilde{\mathcal{H}}(\mathrm{g})
\stackrel{\mathrm{n.r.s.}\,\,}{{-\!\!-\!\!\!}\longrightarrow} 
\widetilde{\mathcal{H}}_{0}\,\,\, 
\mbox{as $\mathrm{g}\to\infty$}.
\label{eq:proof-3}
\end{equation} 
We recall the unitary transformation given 
in Eq.(\ref{eq:unitary-trans-GQR-Hamiltonian}), 
and replace the parameters $\omega_{\mathrm{c}}$ and 
$\mathrm{g}$ with the parameters $\omega_{\mathrm{g}}$ and 
$\widetilde{\mathrm{g}}$ in it, respectively. 
This comes up with the identity,   
\begin{eqnarray}
&{}& 
U(\widetilde{\mathrm{g}}/\omega_{\mathrm{g}})
\mathcal{H}_{\mbox{\tiny GQR}}(\omega_{\mathrm{g}},\widetilde{\mathrm{g}})
U(\widetilde{\mathrm{g}}/\omega_{\mathrm{g}})^{*}
-\widetilde{H}_{\mathrm{ptn}}(\Delta_{\mathrm{g}})
-\hbar\frac{\omega_{\mathrm{g}}}{2} 
\nonumber \\ 
&=& 
H_{\mathrm{ptn}}(\omega_{\mathrm{g}})
-\frac{\hbar}{2}\varepsilon\sigma_{z}
-\frac{\hbar}{2}\omega_{\mathrm{a}}\Xi_{0}(\mathrm{g})
-\widetilde{H}_{\mathrm{ptn}}(\Delta_{\mathrm{g}})
-\hbar\frac{\omega_{\mathrm{g}}}{2} 
\nonumber \\ 
&=& 
\widetilde{H}_{\mathrm{ptn}}(\omega_{\mathrm{g}})
-\frac{\hbar}{2}\varepsilon\sigma_{z}
-\frac{\hbar}{2}\omega_{\mathrm{a}}\Xi_{0}(\mathrm{g})
-\widetilde{H}_{\mathrm{ptn}}(\Delta_{\mathrm{g}}) 
\nonumber \\ 
&=&
\widetilde{\mathcal{H}}(\mathrm{g})
-\hbar\frac{\widetilde{\mathrm{g}}^{2}}{\omega_{\mathrm{g}}}.
\label{eq:proof-4}
\end{eqnarray}
Applying Eq.(\ref{eq:proof-3}) to the Hamiltonian 
$\widetilde{\mathcal{H}}(\mathrm{g})$ in RHS of 
Eq.(\ref{eq:proof-4}) yields the limit 
\begin{equation}
U(\widetilde{\mathrm{g}}/\omega_{\mathrm{g}})
\mathcal{H}_{\mbox{\tiny GQR}}(\omega_{\mathrm{g}},\widetilde{\mathrm{g}})
U(\widetilde{\mathrm{g}}/\omega_{\mathrm{g}})^{*}
-\widetilde{H}_{\mathrm{ptn}}(\Delta_{\mathrm{g}})
-\hbar\frac{\omega_{\mathrm{g}}}{2}
\stackrel{\mathrm{n.r.s.}\,\,}{{-\!\!-\!\!\!}\longrightarrow} 
\widetilde{\mathcal{H}}_{0}-\frac{\hbar}{4C_{\infty}}\,\,\, 
\mbox{as $\mathrm{g}\to\infty$}.
\label{eq:proof-5}
\end{equation}
Since $\mathcal{H}^{\mathrm{ren}}_{A^{2}}(\varepsilon)
=\mathcal{H}_{\mbox{\tiny GQR}}(\omega_{\mathrm{g}},\widetilde{\mathrm{g}})$ 
by Eq.(\ref{eq:renormalized-Hamiltonian}), 
the renormalized Hamiltonian 
$\mathcal{H}^{\mathrm{ren}}_{A^{2}}(\varepsilon)$
can be well approximated by 
\begin{eqnarray*}
&{}&
U(\widetilde{\mathrm{g}}/\omega_{\mathrm{g}})^{*}
\left\{
\widetilde{\mathcal{H}}_{0}
+\widetilde{H}_{\mathrm{ptn}}(\Delta_{\mathrm{g}})
+\hbar\frac{\omega_{\mathrm{g}}}{2}
\right\}U(\widetilde{\mathrm{g}}/\omega_{\mathrm{g}})
-\frac{\hbar}{4C_{\infty}} \\ 
&=&
U(\widetilde{\mathrm{g}}/\omega_{\mathrm{g}})^{*}
\left\{\mathcal{H}_{\mathrm{atm}}(\varepsilon)
+\widetilde{H}_{\mathrm{ptn}}(\omega_{\mathrm{c}}+\Delta_{\mathrm{g}})
+\hbar\frac{\omega_{\mathrm{g}}}{2}
\right\}U(\widetilde{\mathrm{g}}/\omega_{\mathrm{g}})
-\frac{\hbar}{4C_{\infty}} \\ 
&\approx&
U(\widetilde{\mathrm{g}}/\omega_{\mathrm{g}})^{*}
\Bigl\{
\mathcal{H}_{\mathrm{atm}}(\varepsilon)
+\widetilde{H}_{\mathrm{ptn}}(\omega_{\mathrm{g}})
+\hbar\frac{\omega_{\mathrm{g}}}{2}
\Bigr\}U(\widetilde{\mathrm{g}}/\omega_{\mathrm{g}})
-\hbar\frac{\widetilde{\mathrm{g}}^{2}}{\omega_{\mathrm{g}}} \\ 
&=& 
U(\widetilde{\mathrm{g}}/\omega_{\mathrm{g}})^{*}
\Bigl\{
\mathcal{H}_{\mathrm{atm}}(\varepsilon)
+H_{\mathrm{ptn}}(\omega_{\mathrm{g}})
\Bigr\}U(\widetilde{\mathrm{g}}/\omega_{\mathrm{g}})
-\hbar\frac{\widetilde{\mathrm{g}}^{2}}{\omega_{\mathrm{g}}}, 
\end{eqnarray*}
that is, by the limit Hamiltonian 
as in Eq.(\ref{eq:main-result}). 
Here, we used $\widetilde{H}_{\mathrm{ptn}}(\omega_{\mathrm{c}})
+\widetilde{H}_{\mathrm{ptn}}(\Delta_{\mathrm{g}})
=\widetilde{H}_{\mathrm{ptn}}(\omega_{\mathrm{c}}+\Delta_{\mathrm{g}})$, 
and approximations, 
$\omega_{\mathrm{g}}\approx\omega_{\mathrm{c}}+\Delta_{\mathrm{g}}$ 
and 
$\hbar/4C_{\infty}\approx 
\hbar\widetilde{\mathrm{g}}^{2}/\omega_{\mathrm{g}}$, 
respectively secured by Eqs.(\ref{eq:app_omega-g}) 
and (\ref{eq:limit_of_self-energy}).

\section{Conclusion and Discussion}
\label{sec:CD}
We have considered a mathematical establishment of 
the adiabatic approximation 
for the generalized quantum Rabi Hamiltonian 
both without and with the $A^{2}$-term. 
In the case without the $A^{2}$-term, 
we have shown in the adiabatic approximation 
that whether each bare eigenstate 
forms a Schr\"{o}dinger-cat-like entangled state or not 
depends on whether the energy bias in the atom Hamiltonian 
is zero or non-zero. 
On the other hand, in the case with the $A^{2}$-term, 
we have renormalized the $A^{2}$-term 
by employing (meson) pair theory, 
and then, we mathematically established the adiabatic 
approximation for the renormalized Hamiltonian.  
Moreover, we have shown in the adiabatic approximation 
that the Schr\"{o}dinger-cat-likeness appears 
in the both cases where the energy bias is zero and 
where it is non-zero. 

At the end of this section, 
we explain the reason why we take the interest in 
Eqs.(\ref{eq:equivalence1}) and (\ref{eq:equivalence2}). 
We showed in Ref.\cite{HMS} that 
if the $A^{2}$-term effect is sufficiently small, 
then the renormalized Hamiltonian of the generalized quantum Rabi model 
has the chance to have some dressed photons (real photons) 
in the ground state. 
Based on the adiabatic approximation given by 
Eqs.(\ref{eq:approximation_1}) and (\ref{eq:approximation_2}), 
the approximated ground-state expectation 
$N_{0}^{\mathrm{app}}$ 
can be calculated as
$N_{0}^{\mathrm{app}}=\widetilde{\mathrm{g}}^{2}/\omega_{\mathrm{g}}^{2}$ 
as in Eq.(\ref{eq:adiabatic-approximation_N_0}). 
Therefore, each of the adiabatically approximated 
eigenstates has the expression as 
$$
|\mathcal{E}_{n}^{\mathrm{app},\pm}(0)\rangle 
=\frac{1}{\sqrt{2}}
\Bigg(|\!\!\uparrow\rangle 
D\left(-\sqrt{N_{0}^{\mathrm{app}}}\right)
|n\rangle
\pm |\!\!\downarrow\rangle 
D\left(+\sqrt{N_{0}^{\mathrm{app}}}\right)
|n\rangle
\Biggr)
$$ 
for $\varepsilon=0$ 
and 
$$
|\mathcal{E}_{n}^{\mathrm{app},\pm}(\varepsilon)\rangle 
=c_{\pm\varepsilon,\omega_{\mathrm{a}}}
\Biggl(
-\omega_{\mathrm{a}}|\!\!\uparrow\rangle 
D(-\sqrt{N_{0}^{\mathrm{app}}})|n\rangle 
+(\varepsilon\mp\sqrt{\varepsilon^{2}+\omega_{\mathrm{a}}^{2}})
|\!\!\downarrow\rangle 
D(+\sqrt{N_{0}^{\mathrm{app}}})|n\rangle
\Biggr)
$$
for $\varepsilon\ne 0$. 
We note that the approximated ground-state expectation 
$N_{0}^{\mathrm{app}}$ 
can be expressed as 
$N_{0}^{\mathrm{app}}
=
\langle\mathcal{E}_{0}^{\mathrm{app},-}(\varepsilon)|
a^{\dagger}a|\mathcal{E}_{0}^{\mathrm{app},-}(\varepsilon)\rangle$, 
and that the state $|\mathcal{E}_{0}^{\mathrm{app},-}(\varepsilon)\rangle$ is 
the $1$st excited state for sufficiently small $|\varepsilon|$ 
by Eq.(\ref{eq:energy-difference}).  
Therefore, whether the Schr\"{o}dinger-cat-likeness for 
each eigenstate can be observed depends on 
the number of dressed photons in 
the ground state or the $1$st excited state. 
In a sense, namely, 
the `size' of the Schr\"{o}dinger-cat-likeness is determined 
by the number of dressed photons in the ground state. 
This is the reason why we are interested in 
Eqs.(\ref{eq:equivalence1}) and (\ref{eq:equivalence2}).

\qquad 
 
\qquad 

\noindent
\textbf{Acknowledgement} 
This work was supported by JSPS KAKENHI Grant Number JP26310210. 
The author thanks Tomoko Fuse, Kouichi Semba, and Fumiki Yoshihara 
for the valuable discussions on their experimental results, and 
Qing-Hu Chen, Cristiano Ciuti, and Elinor Irish for their 
useful comments at the International Workshop on 
``Strongly Coupled Light-Matter Interactions: Models and Application'' 
organized by Hong-Gang Luo and Jun-Hong An 
at Lanzhou University on July 8 and 9, 2018. 
He also acknowledges Franco Nori's advice on reference literatures.


\newpage 

\begin{thebibliography}{11}



\bibitem{PRSZ} 
Povh~B, Rith~K, Scholz~C and Zetsche~F 2004 
{\it Particles and Nuclei: An Introduction to the Physical Concepts}
(Berlin: Springer)

\bibitem{feynman}
Feynman~R 1985 
{\it QED: The Strange Theory of Light and Matter}
(Princeton: Princeton University Press)

\bibitem{BD1}
Bjorken~J~D and Drell~S~D 1964 
{\it Relativistic Quantum Mechanics} 
(New York: McGraw-Hill Book Company)

\bibitem{BD2}
Bjorken~J~D and Drell~S~D 1965 
{\it Relativistic Quantum Fields} 
(New York: McGraw-Hill Book Company)

\bibitem{cohen-tannoudji}
Cohen-Tannoudji~C, Dupont-Roc~J and Grynberg~G 1992 
{\it Atom-Photon Interaction: Basic Processes and Applications}
(New York: John Willey \& Sons, Inc)

\bibitem{kaku}
Kaku~M 1993 
{\it Quantum Field Theory: A Modern Introduction}
(New York: Oxford University Press)

\bibitem{PS}
Peskin~M~E and Schroeder~D~V 1995
{\it An Introduction to Quantum Field Theory}
(Abingdon: Westview Press)

\bibitem{nori12} 
Nation~P~D, Johansson~J~R, Blencowe~M~P and Nori~F 2012 
Rev. Mod. Phys. \textbf{84}, 1

\bibitem{HT}
Henley~E and Thirring~W 1962 
{\it Elementary Quantum Field Theory} 
(New York: McGraw-Hill)

\bibitem{yukawa}
Yukawa~H 1935 
{\it Proc. Phys.-Math. Soc. Japan} {\bf 17} 48

\bibitem{wentzel1}
Wentzel~G 1941 
{\it Z. Phys.} {\bf 118} 277 

\bibitem{wentzel2}
Wentzel~G 1942 
{\it Helv. Phys. Acta} {\bf 15} 111 

\bibitem{BR}
Brown~L~M and Rechenberg~H 1996 
{\it The Origin of the Concept of Nuclear Forces}
(Bristol: IOP Publishing)

\bibitem{BN} 
Buluta~I and Nori~F 2009 
{\it Science} \textbf{326}, 108

\bibitem{HR-RBH}
Haroche~S and Raimond~J~M 2008, 
{\it Exploring Quantum. Atoms, Cavities, 
and Photons} 
(Oxford: Oxford University Press) 2008,

\bibitem{rbh01}
Raimond~J~M, Brune~M and Harohe~S 2001 
{\it Rev. Mod. Phys.} \textbf{73} 565 

\bibitem{MSS}  
Makhilin~Yu, Sch\"{o}n~G and Shnirman~A 2001 
{\it Rev. Mod. Phys.} \textbf{73} 357

\bibitem{MB} 
Marquardt~F and Bruder~C 2001 
{\it Phys. Rev. B} \textbf{63} 054514

\bibitem{Chiorescu}
Chiorescu~I, Bertet~P, Semba~K, 
Nakamura~Y, Harmans~C~J~P~M and Mooij~J~E 2004 
{\it Nature} \textbf{431} 159--162

\bibitem{Wallraff08} 
Fink~J~M, G\"{o}ppl~M, Bau~M, 
Bianchetti~R, Leek~P~J, Blais~A and Wallraff~A, 2008 
{\it Nature} \textbf{454} 315

\bibitem{Wallraff04} 
Wallraff~A, Schuster~D~I, Blais~A, 
Fruzio~L, Huang~R~-S, Majer~J, 
Kuar~S, Girvin~S~M and Schoelkopf~R~J 2004 
{\it Nature} \textbf{431} 162

\bibitem{YN} 
You~J~Q and Nori~F 2011  
{\it Nature} \textbf{474}, 589

\bibitem{nori17}
Gu~X, Kockum~A~F, Miranowicz~A, Liu~Y~X and Nori~F 2017 
{\it Phys. Rep.} \textbf{718-719}, 1

\bibitem{DGS} 
Devoret~M, Girvin~S and Schoelkopf~R 2007 
{\it Ann. Phys.} \textbf{16} 767

\bibitem{ciuti} 
G\"{u}nter~G, Anappara~A~A, Hees~J, 
Sell~A, Biasiol~G, Sorba~L, 
De Liberato~S, Ciuiti~C, 
Tredicucci~A, Leitenstorfer~A and Huber~R 2009 
{\it Nature} \textbf{458} 178

\bibitem{Mooij} 
Forn-D\'{i}az~P, Lisenfeld~J, Marcos~D, 
Garc\'{i}a-Ripoll~J~J, Solano~E, 
Harmans~C~J~P~M and Mooij~J~E 2010 
{\it Phys. Rev. Lett.} \textbf{105} 237001

\bibitem{Gross} 
Niemczyk~T, Deppe~F, Huebl~H, Menzel~E~P, 
Hocke~F, Schwarz~M~J, Garcia-Ripoll~J~J, 
Zueco~D, H\"{u}mmer~T, Solano~E, 
Marx~A and Gross~R 2010 
{\it Nature Physics} \textbf{6} 772 

\bibitem{FD}
Forn-D\'{i}az~P,  J. J. Garc\'{i}a-Ripoll~J~J, 
Peropadre~B, Orgiazzi~J~-L, Yurtalan~M~A, 
Belyansky~R, Wilson~C~M and Lupascu~A 2017 
{\it Nature Phys.} {\bf 13} 39 

\bibitem{yoshihara1}
Yoshihara~F, Fuse~T, Ashhab~S, Kakuyanagi~K, 
Saito~S and Semba~K 2017 
{\it Nature Phys.} {\bf 13} 44 

\bibitem{yoshihara2}
Yoshihara~F, Fuse~T, Ashhab~S, Kakuyanagi~K, 
Saito~S and Semba~K 2017 
{\it Phys. Rev. A} {\bf 95} 053824

\bibitem{yoshihara3}
Yoshihara~F, Fuse~T, Ao~Z, Ashhab~S, Kakuyanagi~K, 
Saito~S, Aoki~T, Koshino~K and Semba~K 2018 
{\it Phys. Rev. Lett.} {\bf 120} 183601

\bibitem{CS}
Casanova~J, Romero~G, Lizuain~I, Gar\'{c}ia-Ripoll~J~J 
and Solano~E 2010 
{\it Phys. Rev. Lett.} {\bf 105} 263603 

\bibitem{braak1}
Braak D 2011 {\it Phys. Rev. Lett.} {\bf 107} 100401 

\bibitem{AN}
Ashhab~S and Nori~F 2010 {\it Phys. Rev. A} \textbf{81} 042311

\bibitem{irish}
Irish~E~K 2007 {\it Phys. Rev. Lett.} \textbf{99} 173601

\bibitem{fuse}
Xiao~Z, Fuse~T, Ashhab~S, Yoshihara~F, Semba~K, 
Sasaki~M, Takeoka~M and Dowling~J~P, 2018 
arXiv:1807.04927

\bibitem{HMS}
Hirokawa~M, M{\o}ller~J~S and Sasaki~I 
2017 {\it J. Phys A: Math. Theo.} {\bf 50} 184003

\bibitem{RS1}
Reed~M and Simon~B 1980 {\it Method of Modern Mathematical Physics I. Functional Analysis} (San Diego: Academic Press) 

\bibitem{weidmann}
Weidmann~J 1980 {\it Linear Operators in Hilbert Spaces} (New York: Springer-Verlag) 

\bibitem{HH}
Hirokawa~M and Hiroshima~F 2014 {\it Comm. Stoch. Anal.} {\bf 8} 551 

\bibitem{hirokawa15}
Hirokawa~M 2015 {\it Quantum Studies: Math. Found.} {\bf 2} 379

\bibitem{van-hove}
van Hove~L 1952 {\it Physica} {\bf 18} 145

\bibitem{KM}
Klein~A and McCormick~B~H 1955 
{\it Phys. Rev.} {\bf 98} 1428 



\end{thebibliography}
\end{document}